\pdfoutput=1
\documentclass[reprint,amsmath,amssymb,aps,prd,nofootinbib,floatfix,preprintnumbers,superscriptaddress,
]{revtex4-2}
\usepackage{graphicx}
\usepackage{subcaption}
\usepackage{float}
\usepackage{hyperref}
\usepackage{array}
\newcommand{\mc}{\mathcal}

\newcommand{\mr}{\rm}
\newcommand{\br}{\bar{r}}
\newcommand{\bw}{\bar{\omega}}
\hypersetup{
    colorlinks = true,
    linkcolor = blue,
    citecolor = blue,
    urlcolor  = blue
}
\usepackage{mathrsfs}
\begin{document}

\title{Dynamical tidal response of neutron stars via scattering amplitudes}

\author{M.V.S. Saketh}
\email{venkata.saketh@icts.res.in}
\affiliation{International Centre for Theoretical Sciences, Tata Institute of Fundamental Research, Bangalore 560089, India}

\author{Suprovo Ghosh}
\email{S.Ghosh@soton.ac.uk }
\affiliation{Mathematical Sciences and STAG Research Centre, University of Southampton, Southampton, United Kingdom.}

\author{Nils Andersson}
\email{N.A.Andersson@southampton.ac.uk}
\affiliation{Mathematical Sciences and STAG Research Centre, University of Southampton, Southampton, United Kingdom.}
% \affiliation{Your Institution}

%\date{November 2025}
\begin{abstract}
A key challenge of gravitational-wave physics is distinguishing the nature of compact objects in binary coalescences, in particular whether they are black holes or neutron stars. Neutron stars are set apart by a stronger tidal response, whose static and dynamical aspects are directly linked to their rich internal physics. Measurements of this response through gravitational-wave observations constrain the neutron-star equation of state and provide insight into the physics of high-density matter. However, defining the tidal response in general relativity is difficult due to coordinate ambiguities and the complexity of connecting the star's response to the binary dynamics and the associated waveforms. In this paper, we show how the dynamical tidal response of a neutron star can be systematically defined within the worldline effective field theory (EFT) framework, and relate it to the gauge-invariant amplitude for gravitational-wave scattering off an isolated star. We compute this amplitude both within the EFT, using standard quantum field-theory techniques, and within stellar perturbation theory (the corresponding ultraviolet theory), solving the coupled metric and matter perturbation equations numerically in the stellar interior and matching to the analytical Mano-Suzuki-Takasugi (MST) solutions in the vacuum exterior. Matching the amplitude between the two theories fixes the dynamical tidal response. The result is consistent with known expectations, including the static limit and the behaviour near the star's resonant modes, and it recovers the imaginary part of the dominant oscillation mode induced by gravitational-wave dissipation. We conclude with a discussion of potential improvements within both the EFT and perturbation theory.
\end{abstract}

\maketitle

\section{Introduction}
Tidal dynamics induced in
compact binary systems involving neutron stars provide opportunities to probe the nature of matter under extreme conditions. Although demonstrated already in the celebrated case of the first observed neutron-star-binary gravitational-wave signal GW170817~\cite{LIGO_Properties2017,PhysRevLett.119.161101,DES:2017kbs}, the full potential of the strategy is expected to require the development of a new generation of instruments (Cosmic Explorer~\cite{Reitze:2019iox} in the USA and the Einstein Telescope~\cite{ET:2025xjr} in Europe) in the next decade. 

The tidal signature becomes prominent during the late stages of binary inspiral. It has two main contributions. The \emph{static tide}, represented by the so-called Love numbers~\cite{Hinderer_2008} and the associated tidal deformability (the quantity that is inferred from the gravitational-wave signal), was already constrained by the GW170817 data~\cite{LIGOScientific:2018cki}. However, the static tide only relies on bulk properties, like the star's mass and radius. In contrast, the \emph{dynamical tide}---commonly represented by resonances associated with the star's oscillation modes~\cite{Lai:1993di}, in turn, linking the problem to asteroseismology---is sensitive to the specifics of the interior physics. The main dynamical tide contribution---associated with the star's fundamental oscillation mode and already included in state-of-the-art waveform models~\cite{Hinderer2016EffectsON,NRttidalv3} used in the analysis of signals---only weakly so, but the presence (or absence) of lower frequency resonances depends directly on the composition and state of matter. The potential detectability of such features is a high-profile question with immediate relevance for nuclear physics~\cite{fmode_pratten,Counsell_2025,Yu_2017, Gittins:2026ntx}. 

Given the direct link to fundamental physics, there is an ongoing drive to model tidal dynamics to the precision required to meet the demands of future observations. This endeavor requires the development of fully relativistic models. However, efforts in this direction have long been hampered by technical issues. A key challenge is to define the tidal response in a manner that is amenable to quantifying the influence on the binary dynamics and the gravitational-wave signal. This is rendered difficult due to the nonlinearities and coordinate freedom of general relativity. While there has been important progress in recent years in defining the dynamical tidal response of a neutron star~\cite{HegadeKR:2024agt,HegadeKR:2026iou,HegadeKR:2026kku,Andersson:2025iyd}, it remains unclear exactly how existing post-Newtonian (PN) Hamiltonians should be systematically augmented to incorporate the results\footnote{See, however, Ref.~\cite{Racine:2004hs} which derives the equations of motion for systems of arbitrarily structured bodies in general relativity.}. This issue motivates the development of effective field theory (EFT) models~\cite{Levi:2018nxp,Porto:2016pyg,Goldberger2022EffectiveFT}. The last two decades have witnessed the rise of EFTs applied to the two-body problem in general relativity, including the incorporation of spin~\cite{Levi:2015msa,Liu:2021zxr,Cho:2022syn,Marsat:2014xea}, and other finite-size effects~\cite{Goldberger:2004jt,Mandal2023RenormalizingLT} in a systematic PN expansion. The success of these approaches relies on the fact that various theoretical advances developed by particle physicists for computing Feynman diagrams/amplitudes in quantum field theory can now (with some modifications) be applied to the binary problem. This has spawned a host of other approaches (post-Minkowskian, heavy-particle effective theory, etc.)~\cite{PhysRevD.94.104015,Kalin:2020mvi,Aoude:2020onz} involving the application of QFT techniques to gravitational-wave astrophysics. 

The aim of this paper is to provide a robust definition and formulation of the dynamical tidal response within the worldline effective field theory (WEFT) framework, where a compact object is represented as a point particle moving along a worldline, supplemented by additional worldline degrees of freedom that encode finite-size effects such as tides. In this context, we contribute to recent demonstrations~\cite{HegadeKR:2024agt,Andersson:2025iyd} that the frequency-dependent tidal response can be obtained by matching solutions near the stellar surface. We improve upon previous approaches by establishing a direct connection to WEFT while avoiding complications associated with gauge freedom through a mapping between the tidal response and the gravitational Raman/Compton\footnote{We will refer to this as Raman scattering going forward, following the precedent set by Ref.~\cite{Ivanov:2024sds}.} scattering amplitudes; the  process of gravitational waves scattering off a relativistic compact object. We obtain the latter in relativistic stellar perturbation theory which represents the ultraviolet (UV) theory\footnote{Stellar perturbation theory describes physics at high-frequencies and small length scales. Thus, we refer to it as the UV theory in this work.}, using analytical Mano-Sasaki-Takasugi (MST) solutions in the stellar exterior, and numerical solutions in the stellar interior. The tidal response follows by matching the two solutions.  This construction neatly avoids issues of gauge invariance by providing a direct correspondence between observables. 

The method has already been successfully applied to obtain the low-frequency tidal response of black holes. Specifically, the static Love number~\cite{Ivanov:2022qqt}, dissipation numbers at leading and next-to-leading order in frequency~\cite{Saketh:2022xjb}, and the running of the dynamical Love numbers~\cite{Saketh:2023bul} have been determined. There has also been significant work involving scalar perturbations~\cite{Ivanov:2024sds} in a black hole background, and the associated tidal response~\cite{Caron-Huot:2025tlq,Kosmopoulos:2025rfj}. A systematic description of the approach of matching scattering amplitudes to fix the tidal response of scalar, electromagnetic, and gravitational perturbations can be found in Ref.~\cite{Ivanov:2026icp}. Very recently, a closed form expression for the scheme-independent part of the dynamical tidal response of a Schwarzschild black hole was provided in Ref~\cite{Solon:2026ubm}.

For neutron stars, the gravitational Raman scattering amplitude has been used to determine the static Love number and leading dissipation number by working consistently in a low-frequency expansion~\cite{Saketh:2024juq}, to linear order in frequency. Subsequently, Ref.~\cite{Jarequi:2026cyp} extended this up to second order in frequency, albeit relying on the validity of the zone-separation scheme proposed in Ref.~\cite{Ivanov:2022qqt}. An off-shell matching at second order in frequency was obtained recently in Ref.~\cite{Apostolidis:2026qsg}.  Here we extend the WEFT results from Ref.~\cite{Saketh:2024juq} to approximately capture the full frequency dependence of the tidal response\footnote{It is important to mention Ref.~\cite{Chakrabarti:2013lua}, which obtained an approximate expression for the dynamical tidal response of a neutron star by combining worldline EFT and stellar perturbation theory. They found resonant behaviour corresponding to normal-mode oscillations, but faced some issues involving divergent intermediate terms and scheme-dependence when matching the EFT with the UV theory.}. We then match with the arbitrary-frequency result for the scattering amplitude obtained in perturbation theory to get an analytical expression for the dynamical/frequency-dependent tidal response.
The success of our approach relies on two simple but important observations. First, the hierarchy of frequency scales in a neutron star makes it possible to construct a worldline EFT that remains valid even in the vicinity of its low-lying resonances while preserving a post-Newtonian/post-Minkowskian expansion. Second, because the dynamical tidal response of a black hole is parametrically suppressed relative to that of a neutron star by powers of the compactness, the difference between the two is dominated by the neutron-star response. We therefore organize the EFT around the black-hole point-particle theory and systematically add the neutron-star tidal response. This organization not only captures dynamical tidal effects efficiently but also mirrors the strategy adopted by many waveform models, which treat binary black holes as the baseline and incorporate neutron-star tidal effects as corrections~\cite{Gralla:2017djj}. We provide evidence that our result offers a more realistic description of dynamical tides by studying the static limit and the behaviour near individual stellar oscillation frequencies. Our work, therefore, brings us closer to a robust understanding of the influence of mode resonances on the orbital evolution of binary neutron-star systems.

The paper is organized as follows:
In Sec.~\ref{secII:SPT}, we briefly review stellar perturbation theory in the interior and exterior of the star. The interior contains coupled metric and matter perturbation equations that will be solved numerically, while the exterior contains only metric perturbations treated analytically using the Mano-Sasaki-Takasugi formalism. Combining these solutions yields the tidal scattering phase for gravitational waves. In Sec.~\ref{secIII:EFT}, we introduce the worldline effective field theory description of a compact object coupled to gravity and compute the corresponding gravitational Raman scattering amplitude. In Sec.~\ref{secIV:RES}, we match the scattering amplitude in EFT to that obtained in perturbation theory to derive an expression for the dynamical tidal response. We then proceed to compare  this representation of the tidal response to results from the literature, demonstrating how the new formalism improves upon previous work. Finally, Sec.~\ref{secV:CONC} concludes with a discussion and suggestions for  future improvements.

We work with units $G=c=\hbar=1$, and every instance of $M$ should be treated as $GM/c^2$, with dimensions of length. We use the mostly positive metric signature $(-,+,+,+)$ throughout the paper.

\section{Stellar perturbation theory: A brief overview}
\label{secII:SPT}

\subsection{Perturbation equations in the interior.}
We consider linear perturbations of a nonrotating, spherically symmetric relativistic star.  The metric tensor that describes the equilibrium background follows from
\begin{equation}
ds^2 = -e^{\nu(r)} dt^2 + e^{\lambda(r)} dr^2 + r^2(d\theta^2+\sin^2\theta\,d\phi^2),
\end{equation}
together with a perfect fluid stress-energy tensor 
\begin{equation}
    T^{\mu\nu}=(\varepsilon+p)u^\mu u^\nu + p g^{\mu\nu},
\end{equation}
where $\varepsilon$ and $p$ represent the energy density and the pressure, respectively.

We introduce linear Eulerian perturbations ($h_{\mu\nu}=g_{\mu\nu}-g_{\mu\nu}^{(0)}$) to the metric, along with a Lagrangian fluid displacement vector $\xi^{\mu}$, perturbations of the fluid 4-velocity ($u^{\mu}$), and the thermodynamical quantities $p$ and $\varepsilon$. Because of spherical symmetry, all perturbations can be decomposed into tensor spherical harmonics with angular indices $(\ell,m)$. Adopting the
Regge-Wheeler gauge and focusing on the polar sector, the metric
perturbations then take the form
\begin{equation}
\begin{split}
    & h_{\mu\nu}^{\text{polar}} =- Y_{\ell m}(\theta,\phi)\, e^{-i\omega t}\times\\
    &\begin{pmatrix}
e^{\nu} r^\ell H_0(r) & i\omega r^{\ell+1} {H}_1(r) & 0 & 0 \\
i\omega r^{\ell+1} {H}_1(r) & e^{\lambda} r^\ell {H}_2(r) & 0 & 0 \\
0 & 0 & r^{\ell+2} {K}(r) & 0 \\
0 & 0 & 0 & r^{\ell+2}\sin^2\theta \,{K}(r)
\end{pmatrix}
.    
\end{split}
\label{eq:polar-r-l}
\end{equation}
Meanwhile, using the gauge-freedom associated with the Lagrangian formulation, we set $u^{\mu}\xi_{\mu} = 0$, and expand the non-trivial components as
\begin{eqnarray}
%\begin{split}
 \xi^r(r) &=& e^{-\lambda/2}r^{\ell - 1}W(r) Y_{\ell m} e^{-i\omega t}, \\
\xi^\theta& =& -r^{\ell - 2}V(r)\partial_\theta Y_{\ell m} e^{-i\omega t},\\
\xi^\phi &=& -\frac{r^{\ell - 2}}{\sin^2{\theta}}V(r)\partial_\phi Y_{\ell m} e^{-i\omega t}.   
%\end{split}
\end{eqnarray}

In summary, a general perturbation is described by the spacetime variables $H_0(r), H_1(r), H_2(r)$ and $K(r)$, alongside the fluid variables $W(r)$ and $V(r)$. These perturbation functions are, however, not all independent. We follow the formalism developed by Detweiler and Lindblom~\cite{Detweiler1985} to reduce the perturbation equations for the interior of the star to a system of four first-order differential equations for $H_1(r), K(r), W(r)$ and $X(r)$ (see Eq. 8-11 in Detweiler and Lindblom~\cite{Detweiler1985}), where $X(r)$ is an auxiliary variable proportional to the Lagrangian pressure perturbation, defined as
\begin{multline}
%\begin{split}
        X(r) = \omega^3(p+\rho)e^{-\nu/2}V - r^{-1}p'e^{(\nu-\lambda)/2}W\\  + \frac{1}{2}(p+\rho) e^{\nu/2}H_0 .
%\end{split}
\end{multline}

The four independent perturbation equations are solved subject to boundary conditions. At the center of the star, at $r = 0$, the solutions must be regular. The appropriate solutions are determined via Taylor series expansion as described in Section II of Detweiler and Lindblom~\cite{Detweiler1985}. Outside the stellar surface, where the fluid perturbations vanish, we only have to solve for perturbations of the spacetime. In essence, the internal perturbations are matched to the standard Regge-Wheeler variable outside the star, as described in Sec.~\ref{sec:PetrubOutside}.

To close the fluid perturbation problem inside the star, we need to provide a matter equation of state; including  a relation between perturbed (Lagrangian) pressure ($\Delta P$) and energy density ($\Delta \varepsilon$). This relation is generally expressed in terms of the sound speed of the dense matter
\begin{equation}
\label{eq:soundspeed}
    \Delta P = \left(\frac{\partial P}{\partial \varepsilon}\right)_{n_i,T}\Delta \varepsilon = c_{ad}^2\Delta \varepsilon,
\end{equation}
where $n_i$ are the densities of the constituent particles and $T$ is the temperature. Since we are interested in the inspiral phase of a compact binary system, the temperature of the star is expected to be $\ll 1$ MeV~\cite{Lai:1993di,Ghosh:2023vrx}, and can be neglected for all practical purposes. Now, the exact expression for the sound speed depends on the underlying thermodynamical conditions, the state and composition of matter. In this work, we are mainly interested in $npe\mu$ matter, which maintains the background chemical equilibrium via the Urca reactions. At low temperatures, the reaction rates are slow enough (compared to the perturbation timescales) to maintain $\beta$-equilibrium~\cite{Sawyer1989}. Hence, the matter composition is practically frozen. In this scenario, the sound speed is given Eq.~\eqref{eq:soundspeed}.%, which can be very different from the equilibrium sound speed $c^2_{eq} = \left(\frac{dP}{d\varepsilon}\right)$. 
If, on the other hand, the reactions were to be fast (compared to the perturbation timescales), the perturbed pressure would maintain the same relationship as the equilibrium background, i.e. we would have
\begin{equation}
    \Delta P  = c_{eq}^2\Delta \varepsilon = \left(\frac{dP}{d\varepsilon}\right) \Delta \varepsilon.
\end{equation}
For a realistic neutron star model, the two sound speeds can be very different, and the difference $c_{ad}^2 - c_{eq}^2$ then gives rise to $g$-modes~\cite{Counsell:2023pqp}. Given that the typical frequencies of low-order $g-$modes are $\sim 100 - 500$ Hz~\cite{Counsell:2024pua,Tran:2022dva}, depending on the mass and interior composition of the neutron star, they can be resonantly excited by the tidal driving during the late stages of binary inspiral. However, at the frequencies relevant for higher order $g$-modes, the standard perturbation formulation meets with numerical difficulties~\cite{Kruger_2015}. To alleviate this, we adopt the strategy described in Appendix B of Kruger et al~\cite{Kruger_2015} (essentially using $V$ instead of $X$ as an independent variable).

\subsection{Perturbation equations in the exterior}
\label{sec:PetrubOutside}

In the stellar exterior, the metric perturbation equations can be mapped to a single second-order differential equation, the Regge-Wheeler (RW) equation~\cite{PhysRev.108.1063}. For a given mode, (i.e., $\ell,m,\omega$) the RW equation is
\begin{equation}
\frac{d^2\phi}{dr_*^2}+[\omega^2-V(r)]\phi=0,
\end{equation}
with
\begin{equation}
V(r) = f(r)\left[\frac{3 f(r)}{r^2}+\frac{(\ell^2+\ell-3)}{r^2}\right],
\end{equation}
 the tortoise coordinate 
\begin{equation}
   r_* = r+2M\log\left(\frac{r}{2M}-1\right) ,
\end{equation}
 and $f(r)=1-(2M/r)$.

Typically, one maps axial metric perturbations to the RW equation, while the polar perturbations considered here are mapped to the Zerilli equation~\cite{Zerilli:1970se}. However, as the two sets of vacuum perturbations can be mapped to one another, we have chosen to work with the RW equation (which has a slightly simpler potential).

\begin{widetext}
Specifically, the RW scalar is related to the metric perturbations in the polar sector through the relations
\begin{equation}
        \phi(r) =   \frac{2M(1+n)f(r)(3M+nr)}{M(1+n)-r^3\omega^2}H_0(r) ~+  \frac{r(n(1+n)r-3M(n+n^2+r^2\omega^2))}{M(1+n)-r^3\omega^2}K(r),\label{eq:phi(H,K)}
\end{equation}
\begin{multline}
 \frac{d\phi(r)}{dr} = \frac{(1+n)(6M^2-n(1+n)r^2)}{r^4\omega^2-M(1+n)}H_0(r)+ \frac{3Mr(-1-n+r^2\omega^2)}{r^4\omega^2-M(1+n)}H_0(r) \\ + \frac{n r(-3M+r(1+n))(1+n-r^2\omega^2)}{2M f(r) (M(1+n) -r^3\omega^2)} K(r)  + \frac{(3M-(n+1)r)(2n(1+n)+3r^2\omega^2)}{2f(r)(M(1+n)-r^3\omega^2)}K(r),
    \label{eq:dphi(H,K)}
\end{multline}
where $n=(l-1)(l+2)/2$.

Moreover, the RW equation has analytical Mano-Sasaki-Takasugi (MST) solutions~\cite{Mano:1996mf}, given by sums over hypergeometric functions. A detailed discussion of the MST solutions for the RW equation can be found in Refs.~\cite{Casals:2015nja}. Here, we only need some of the results from that paper.

In general, as a second-order differential equation, the RW equation has two independent solutions, whose coefficients are determined by the boundary conditions at stellar surface. A convenient basis of solutions is given by
\begin{multline}
        X_0^\nu  =(1-x)^{\nu+1+i\bw}e^{i\bw x}(-x)^{-i\bw}\sum_{n=-\infty}^{n=\infty}\frac{\Gamma(-n-\nu+s-i\bw)\Gamma(2n+2\nu+1)}{\Gamma(n+\nu+1-s-i\bw)}a_n (1-x)^n \\  \times {}_2F_1(-n-\nu+s -i\bw, -n-\nu-s-i\bw, -2n-2\nu;\br^{-1}), 
            \label{eq:near_zone_basis1}
\end{multline}
\begin{equation}
        X_0^{-\nu-1} = X_0^{\nu}|_{\nu=-\nu-1}.
                    \label{eq:near_zone_basis2}
\end{equation}
\end{widetext}
Here, $a_n$ are coefficients which satisfy a recursion relation, and $\nu$ is called the renormalized angular momentum which is equal to $\ell$ in the limit $\omega\rightarrow 0$. Other variables are scaled in such a way that $\bar{\omega}=2M\omega$, $\bar{r}=r/(2M)$ and $x=1-\bar{r}$. $s$ is the spin weight of external perturbations, i.e., we have  $s=-2$ for gravitational perturbations.  Finally, ${}_2F_1$ denotes the Gauss hypergeometric function.

As given in  Eq.~\eqref{eq:near_zone_basis1} and \eqref{eq:near_zone_basis2}, the expansions converges and are well-defined in the region $r\in(2M,\infty)$. However, for a given frequency $\omega$, their convergence radius is limited from above by $r\simeq \omega^{-1}$. Thus, the general RW scalar is given by
\begin{equation}
 %   \begin{split}
        \phi = B_{\nu}X_0^\nu + B_{-\nu-1}X_0^{-\nu-1}.
%    \end{split}
\end{equation}
The frequency-dependent ratio $B_{-\nu-1}/B_{\nu}$ then encodes the rich internal physics of a neutron star (it is equal to 1 for black holes). For a neutron star, the relevant value is obtained by matching to the metric perturbations at the stellar surface (at $r=R$) using Eq.~\eqref{eq:phi(H,K)} and Eq.~\eqref{eq:dphi(H,K)}. This determines the solution in the stellar exterior. %, and we have implemented the computation of $B_{-\nu-1}/B_{\nu}$ from the boundary conditions at stellar surface in Mathematica.

\subsection{Gravitational Raman scattering phase}

To compute the scattering phase, we need to study the behaviour of $\phi(r)$ in the wave zone, i.e. in the limit $r\gg \omega^{-1}$.  However, as mentioned earlier, the expansions in \eqref{eq:near_zone_basis1} and \eqref{eq:near_zone_basis2} do not converge in the regime $\omega r\gg 1$. In order to circumvent this issue, we note that the solutions themselves remain valid, and can be related to another basis of solutions which have well-defined expansions in the far-zone. 

The new solutions, denoted $X_C^{\nu/-\nu-1}$,  are related to $X_0^{\nu/-\nu-1}$ according to
\begin{eqnarray}
    X_0^\nu &=& K_\nu X_C^\nu \\
    X_0^{-\nu-1}&=&K_{-\nu-1}X_C^{-\nu-1},
\end{eqnarray}
where the expression for $K_\nu$ can be found in Eq.~(3.32) of Ref.~\cite{Casals:2015nja}.\footnote{ There is a small typo in  Eq.~(3.32) of Ref.~\cite{Casals:2015nja}. In the numerator, it should be $(2\bar{\omega})^{s-\nu-r-1}$, and not $(2\bar{\omega})^{s-\nu-r}$.} The expansions for $X_C^\nu$ can also be found in Eq.~(3.26) of Ref.~\cite{Casals:2015nja}.

Thus, in the wave-zone, we can instead work with
\begin{equation}
    \phi(r) = B_{\nu}K_{\nu}X_C^\nu + B_{-\nu-1}K_{-\nu-1}X_C^{-\nu-1}.
\end{equation}
In the limit $\omega r\gg 1$, this takes the form
\begin{equation}
        \phi_{\ell \omega}(r)|_{r\rightarrow \infty} = A_{\ell \omega}^{\rm out}e^{i\omega r} + A_{\ell \omega}^{\rm in} e^{-i\omega r},
\end{equation}
with
\begin{multline}
        \frac{A^{\rm out}_{\ell \omega}}{A^{\rm in}_{\ell \omega}} = (-1)^{\ell + 1} e^{2i\delta^{\rm NS}_{\ell \omega}} \\
        =\frac{e^{2i\bw \log\bw}A_-^{\nu}(1+ie^{i\pi \nu}\mc{K}_{\text{NS}})}{A_+^{\nu}\left(1-ie^{-i\pi \nu}\frac{\sin[\pi(\nu+i\bw)]}{\sin[\pi(\nu-i\bw)]}\mc{K}_{\text{NS}}\right)} ,
    \label{eq:SPT_phase}
\end{multline}
and 
\begin{equation}
   \mc{K}_{\mr{NS}}=\frac{B_{-\nu-1}K_{-\nu-1}}{B_{\nu} K_{\nu}} .
\end{equation}
The expressions for $A_{-/+}^\nu$ can be found in Eqs.~(3.19) and (3.41) in Ref.~\cite{Casals:2015nja}. In principle, we now have an expression for the scattering phase $\delta_{\ell \omega}$ as a function of the stellar boundary conditions. While this would be a satisfactory point to conclude the discussion of the perturbation theory, it is convenient to instead subtract the corresponding result for a black hole and define the ``tidal'' phase 
\begin{equation}
    \delta^{\rm tid}_{\ell \omega} = \delta^{\rm NS}_{\ell \omega}-\Re\delta^{\rm BH}_{\ell \omega}.
\end{equation} 
We can obtain $\Re\delta^{\rm BH}_{\ell \omega}$ directly from Eq.~\eqref{eq:SPT_phase} since we know that $B_{-\nu-1}=B_{\nu}$ for a black hole. 

\begin{widetext}
The advantage of this definition is that it removes contributions to the scattering phase that are common to both neutron stars and black holes, namely the non-tidal effects associated with propagation through the Schwarzschild spacetime exterior. This is useful for two reasons. First, from a practical point of view, state-of-the-art gravitational waveform models commonly use the black hole result as a baseline to which additional features, like neutron star tides, are added~\cite{Gralla:2017djj}. Second, from a formal perspective; since black holes have vanishing static Love numbers and a tidal response that is suppressed relative to that of neutron stars, the resulting phase shift $\delta^{\rm tidal}_{\ell\omega}$ is dominated by the neutron-star tidal response. We will make this statement more precise within the EFT framework in Sec.~\ref{secIII:EFT}. An additional benefit is that this definition eliminates the factors appearing in Eq.~\eqref{eq:SPT_phase}, namely $e^{2i\bw\log\bw}A_-^\nu/A_+^\nu$. These encode far-zone physics that is insensitive to the internal structure of the compact object~\cite{Ivanov:2022qqt}. Indeed, they are identical for black holes and neutron stars, as they do not depend on $B_{\nu(-\nu-1)}$.

Explicitly, we can now write
\begin{equation}
%\begin{split}
2i \delta^{\rm tidal}_{\ell \omega} = \log\left[\frac{(-1)^{\ell+1}e^{2i\bw \log(\bw)}A_-^{\nu}(1+ie^{i\pi \nu}\mc{K}_{\text{NS}})}{A_+^{\nu}\left(1-ie^{-i\pi \nu}\frac{\sin[\pi(\nu+i\bw)]}{\sin[\pi(\nu-i\bw)]}\mc{K}_{\text{NS}}\right)}\right] - i\Im\log\left[\frac{(-1)^{\ell+1}e^{2i\bw \log(\bw)}A_-^{\nu}(1+ie^{i\pi \nu}\mc{K}_{\text{BH}})}{A_+^{\nu}\left(1-ie^{-i\pi \nu}\frac{\sin[\pi(\nu+i\bw)]}{\sin[\pi(\nu-i\bw)]}\mc{K}_{\text{BH}}\right)}\right] ,
%\end{split}
\end{equation}
where $\mc{K}_{\rm BH}=K_{-\nu-1}/K_\nu$. Moreover, we can rewrite $(-1)^{\ell+1}e^{2i\bar{\omega}\log\bar{\omega}}(A_-^\nu/A_+^\nu)=e^{2i\delta_{\rm far}}$, where $\delta_{\rm far}$ is purely real and independent of the nature of the compact object. These factors therefore cancel, leaving us with
\begin{equation}
%\begin{split}
2i \delta^{\rm tidal}_{\ell \omega} = \log\left(\frac{1+ie^{i\pi \nu}\mc{K}_{\text{NS}}}{1-ie^{-i\pi \nu}\frac{\sin[\pi(\nu+i\bw)]}{\sin[\pi(\nu-i\bw)]}\mc{K}_{\text{NS}}}\right) - i\Im\log\left(\frac{1+ie^{i\pi \nu}\mc{K}_{\text{BH}}}{1-ie^{-i\pi \nu}\frac{\sin[\pi(\nu+i\bw)]}{\sin[\pi(\nu-i\bw)]}\mc{K}_{\text{BH}}}\right) ,
%\end{split}
\label{eq:SPT_tidal_phase}
\end{equation}

Finally, let us justify the term ``tidal phase.'' In the regime $M\omega\ll1$, the dynamical tidal response of a black hole is negligible compared to that of an neutron star. Moreover, the static Love number of a black hole vanishes, while taking the real part of $\delta_{\rm BH}$ removes dissipative effects associated with horizon absorption. Consequently, $\delta^{\rm tidal}_{\ell\omega}$ may be viewed as the neutron-star phase shift with the universal point-particle contribution removed.
\end{widetext}

While this concludes the scattering problem in stellar perturbation theory, Eq.~\eqref{eq:SPT_tidal_phase} is of limited utility for the binary problem unless we provide the means to extract the tidal response from it. To do this, we need to derive $\delta_{\ell m}^{\rm tidal}$ as a function of the tidal response, after suitably defining the latter. With this in mind, we turn to describing a worldline EFT for a tidally deformed neutron star. The intention is to properly define the tidal response and relate it to the tidal scattering phase.

\section {Worldline effective field theory including dynamical tidal effects}
\label{secIII:EFT}

We aim to model the compact object (a neutron star) as a point-particle following a worldline $z^\mu(\tau)$, adding degrees of freedom to capture the main finite-size effects. Formally, such a description is possible when the wavelength of the perturbations is much larger than the size of the compact object, i.e., when $R\omega\ll 1$ and $M\omega\ll 1$. Within the worldline effective field theory framework, the separation of scales naturally motivates an expansion in the perturbing frequency. In many applications, one assumes that the perturbing frequency $\omega$ is small compared to all intrinsic frequency scales of the system. For realistic neutron stars with dynamical tides, however, this assumption generally breaks down. In particular, near resonance the perturbing frequency becomes comparable to the frequency  of the relevant stellar oscillation modes, i.e., $\omega \sim \omega_{f,g,\dots}$. Given that a stratified neutron star supports a formally infinite spectrum of low-frequency gravity modes the tidal response cannot, in general, be captured by an expansion in $\omega/\omega_{f,g,\dots}$. We therefore adopt the weaker assumption $M\omega \ll 1$, allowing $\omega$ to be comparable to the characteristic mode frequencies. This provides a framework capable of describing dynamical tidal effects, including resonant mode excitations. 

Stated differently, neutron stars exhibit a hierarchy of intrinsic frequency scales, unlike black holes, whose only characteristic dynamical scale is $M^{-1}$. Several of the neutron star's internal scales are parametrically smaller than $M^{-1}$, giving rise to oscillation modes at (significantly) lower frequencies than the black-hole quasinormal modes. This scale separation permits an EFT that remains valid even when the orbital frequency approaches these low-frequency modes, capturing resonant finite-size effects while retaining a post-Newtonian/post-Minkowskian (PN/PM) expansion in $M\omega$. Since the fundamental ($f$-) mode  typically represents the dominant tidal excitation, the EFT should remain applicable at least up to frequencies commensurate with the $f$-mode. For a non-spinning neutron star, $f_f\sim1-3\,\mathrm{kHz}$ (equivalently, $\omega_f\sim(0.02-0.06)\,M_\odot^{-1}$ for a $1.4\,M_\odot$ star), whereas the relativistic scale is $M^{-1}\simeq23\,\mathrm{kHz}$ (or $\omega\sim2\pi\times23\,\mathrm{kHz}$). Thus, the $f$-mode lies more than an order of magnitude below $M^{-1}$, justifying an EFT that is nonperturbative in $\omega/\omega_{g,f,\dots}$ while remaining perturbative in $M\omega$.

Let us, first of all, write down a worldline action for a spinless particle with a quadrupolar degree of freedom, coupled to external tidal fields as\footnote{We are using the Polyakov form of the worldline action with the einbein set to 1.}
\begin{multline}
S = \int d\tau \frac{M}{2}(g_{\mu\nu}\dot{z}^\mu \dot{z}^\nu-1) 
       + \int d\tau \frac{1}{2}Q^{\mu\nu}(\tau)
        E_{\mu\nu}\!\left(z^{\mu}\right)
    \\ + S_Q
       + \frac{1}{16\pi}\int d^4x\,R\sqrt{-g}
      + S_{\rm GF},
   \label{eq:action2}
\end{multline}
or 
\begin{multline}
S = \int d\tau \frac{M}{2}(g_{\mu\nu}\dot{z}^\mu \dot{z}^\nu-1) 
       - \int d\tau\dot{z}^{[\mu}Q^{\nu][\rho}\dot{z}^{\sigma]}
        R_{\mu\nu\rho\sigma}
    \\  + S_Q
       + \frac{1}{16\pi}\int d^4x\,R\sqrt{-g}
      + S_{\rm GF},
\end{multline}
where $\tau$, the worldline parameter is also identified with the proper time. With dots representing differentiation with respect to $\tau$, the electric tidal field is then given by $E_{\mu\nu}(x)=C_{\mu\alpha\nu\beta}\dot{z}^\alpha \dot{z}^\beta$. The proper time $\tau$ is measured by a co-moving, inertial observer sufficiently far away from the compact object, the mass of which is $M$. $Q^{\mu\nu}(\tau)$ is used to model the tidal deformation, which should  ideally be represented in terms of observable quantities. One way to acheive this is to write down an Ansatz for the quadrupole moment, expressing it as a linear functional of the tidal field $E_{\mu\nu}(z^\mu(\tau))$~\cite{Saketh:2023bul}. The tidal response function can then be defined in terms of this linear functional, and may be regarded as the Green's function used to relate the source (the tidal field) to the excitation (the quadrupole moment). An equivalent approach would be to define the tidal response function in terms of the Feynman propagator of the quadrupole moment. This approach is more convenient as it enables the use of Feynman-diagram machinery for computing the Raman scattering amplitude. This follows since the Feynman propagator is related to a Green's function used for computing a field in the presence of a source, and is thus suitable for defining the tidal response. We  also need to impose the constraints $Q^{\mu\nu}\dot{z}_\nu=0$, $Q^{\mu}{}_{\mu}=0$, and $Q^{\mu\nu}=Q^{\nu\mu}$, as a physical quadrupole moment must have only 5 spacelike degrees of freedom. We will impose this constraint when defining the tidal response below. 

The action in Eq.~\eqref{eq:action2} determines the dynamics of the gravitational field in the environment of the compact object. Treating the problem perturbatively, we expand 
the action in terms of deviations from flat spacetime, i.e., $g_{\mu\nu}=\eta_{\mu\nu}+h_{\mu\nu}$. In addition,  we choose to express $S_{\rm GF}$  in harmonic gauge where $\partial_\mu h^{\mu\nu} = (1/2)\partial^\nu h$, with the usual gauge-parameter set to 1.
Since the object's motion may deviate due to external gravitational peturbations, we also expand the worldline as $z^\mu= u^\mu \tau + \delta z^\mu$, where $\delta z^\mu \sim \mc{O}(h)$, where $u^\mu$ is the constant 4-velocity of the unperturbed object. We then obtain

\begin{widetext}
\begin{multline}
 S = \frac{1}{2}\int d\tau \left(-2+ h_{\mu\nu}u^\mu u^\nu + 2 h_{\mu\nu} u^\mu \dot{\delta z}^\nu + \delta z^\alpha u^\mu u^\nu \partial_\alpha h_{\mu\nu}+\eta_{\mu\nu} \dot{\delta z}^\mu \dot{\delta z}^\nu \right)\,M + S_Q
\\ - \int d\tau \, u^{[\mu}Q^{\nu][\rho}u^{\sigma]}
\partial_\rho\partial_\nu h_{\mu \sigma}
 - \frac{1}{32 \pi}\int d^4 x \,
\frac{1}{2}\partial_\lambda h_{\mu\nu}
P^{\mu\nu,\rho\sigma}\partial^\lambda h_{\rho\sigma}
 + \mathcal{O}(h^3),
\end{multline}
or
\begin{multline}
S \approx -\int  M d\tau  + M\int d\tau \frac{\tilde{h}_{\mu\nu}}{2 m_{\rm pl}}u^\mu u^\nu + M \int d\tau  \left(\delta z^\alpha  f_\alpha% (\frac{1}{2}u^\mu u^\nu \partial_\alpha h_{\mu\nu}+ h_{\mu\nu} u^\mu \dot{\delta z}^\nu )
+ \frac{1}{2}\eta_{\mu\nu} \dot{\delta z}^\mu \dot{\delta z}^\nu \right)
+ S_Q
\\ -\frac{1}{m_{\rm pl}}\int d\tau \,
u^{[\mu}Q^{\nu][\rho}u^{\sigma]}
\partial_\rho\partial_\nu \tilde{h}_{\mu \sigma} 
-  \int d^4 x \,
\frac{1}{2}\partial_\lambda \tilde{h}_{\mu\nu}
P^{\mu\nu,\rho\sigma}\partial^\lambda \tilde{h}_{\rho\sigma}
 + \mathcal{O}(h^3),
 \label{eq:action_expanded}
 \end{multline}
 with
 \begin{equation}
 f_\alpha = -\eta_{\alpha\lambda}\Gamma^{\lambda}_{\mu\nu}u^\mu u^\nu = \frac{1}{2}(\partial_\alpha h_{\mu\nu} - 2 \partial_\nu h_{\alpha \mu})u^\mu u^\nu
\end{equation}
and
\begin{equation}
m_{\rm pl}^{-1}\tilde{h}_{\mu\nu}=h_{\mu\nu}, 
\qquad
m_{\rm pl}^2=\frac{1}{32\pi},
\end{equation}
Alternatively, 
\begin{multline}
 S_{\rm eff}=-\int  M d\tau \left(1-\frac{\tilde{h}_{\mu\nu}}{2 m_{\rm pl}}u^\mu u^\nu\right) +  \int d\tau \frac{M}{2}f^\nu \frac{1}{\partial_t^2} f_\nu 
+ S_Q 
\\
-\frac{1}{m_{\rm pl}}\int d\tau \,
u^{[\mu}Q^{\nu][\rho}u^{\sigma]}
\partial_\rho\partial_\nu \tilde{h}_{\mu \sigma} 
 - \int d^4 x \,
\frac{1}{2}\partial_\lambda \tilde{h}_{\mu\nu}
P^{\mu\nu,\rho\sigma}\partial^\lambda \tilde{h}_{\rho\sigma} 
+ \mathcal{O}(h^3),
\end{multline}
where
\begin{equation}
P^{\mu\nu,\rho\sigma}
= \frac{1}{2}
\left(
\eta^{\mu\rho}\eta^{\nu\sigma}
+\eta^{\mu\sigma}\eta^{\nu\rho}
-\eta^{\mu\nu}\eta^{\rho\sigma}
\right).
\end{equation}
Here we have substituted the equation of motion for $\delta z^\mu$ at leading order in the metric perturbation in order to eliminate it. This leaves behind a non-local recoil term ($2^{-1}M\int d\tau f^\nu \partial_t^{-2}f_\nu$), accounting for the backreaction of the compact object in response to the emission of gravitational waves. This term is essential to preserve gauge invariance~\cite{Ivanov:2026icp}. Note, however,  that $f^\mu$ vanishes in traceless-transverse gauge where $h^{\mu\nu}u_\nu=0$.

We assume $Q^{\mu\nu}=0$ for the unperturbed star and take $S_Q$ to be at least quadratic in $Q$, implying that perturbations to $S_Q$ are irrelevant for the computation of the Raman scattering amplitude. Specifically, such perturbations can be either absorbed into redefinitions of the mass or spin of the star~\cite{Saketh:2022xjb,Mandal:2023hqa}, vanish in dimensional regularization, give rise only to trivial divergences (i.e., divergences that can be removed by infinite counterterms without generating running logarithms), or fail to contribute to the Raman scattering process in the classical limit. For the same reason, we need not expand the four-velocity contracted with the quadrupole tensor, and instead replace it by its background value $u^\mu$.

The multi-graviton interactions owing to the nonlinear nature of general relativity are encoded in $\mc{O}(h^3)$ terms that we have not explicitly written out, but which will be relevant in the analysis that follows. Note that we still write the action covariantly, without specializing to $u^\mu\equiv(1,0,0,0)$. This is just a matter of convenience. However, we will label external graviton states based on their momentum and frequency in the frame defined by $u^\mu$.

We now want to study the scattering amplitude of gravitational waves within the WEFT framework. Making use of standard quantum field theory techniques, we can write down the  Feynman rules in momentum space, some of which are shown in Fig.~\ref{fig:Frules} where $\mathscr{P}^{\mu\nu} = \eta^{\mu\nu}+u^{\mu}u^{\nu}$. 
\begin{figure}[h!]
    \centering
    \includegraphics[width=0.6
    \linewidth]{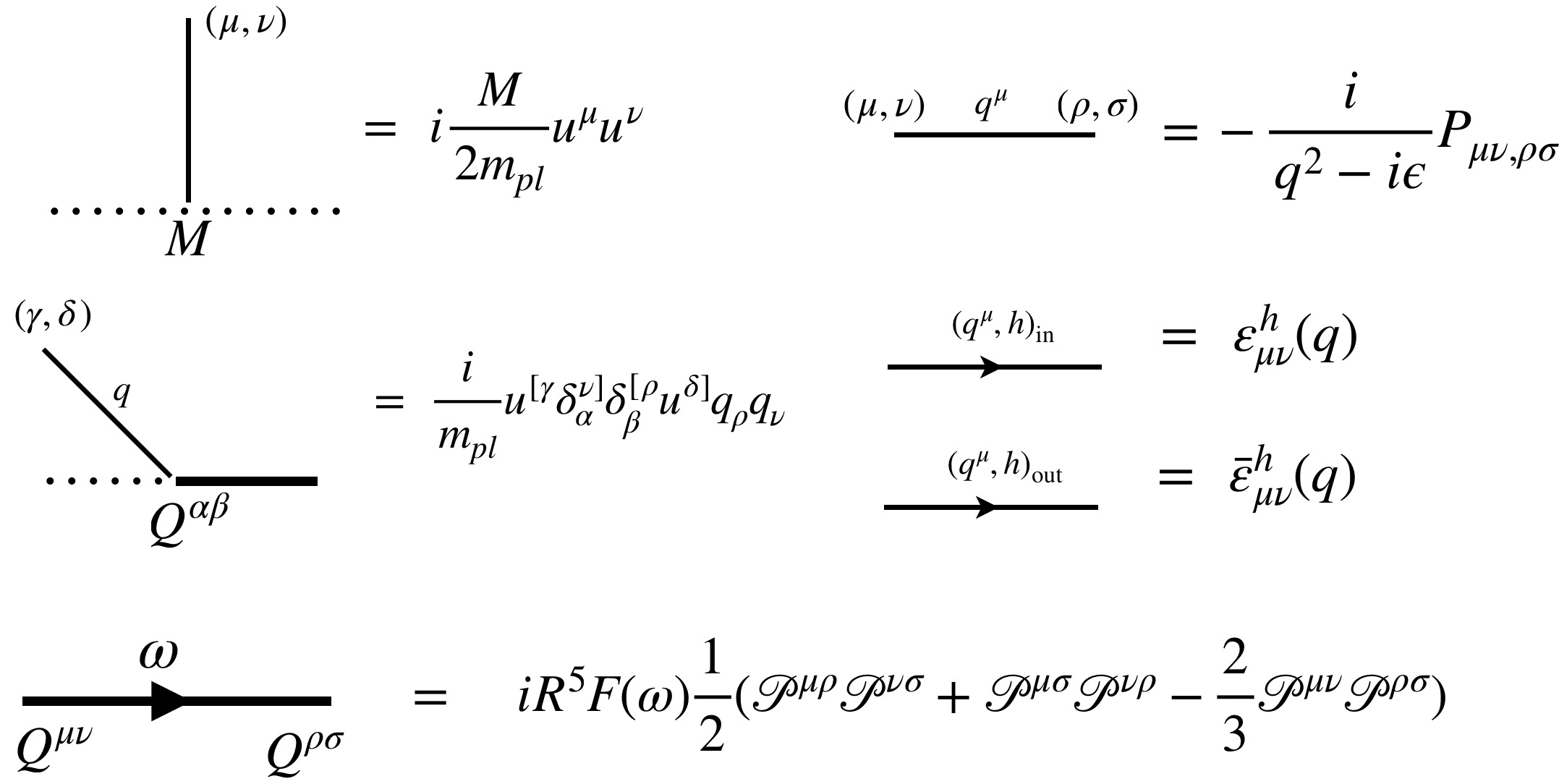}
    \caption{Some of the Feynman rules for WEFT in momentum space.}
    \label{fig:Frules}
\end{figure}
There are additional rules for multi-graviton interaction vertices which we ignore for the moment. We also have the recoil term, the contribution of which to the scattering amplitude vanishes at tree-level in a suitable gauge (e.g., traceless-transverse gauge). Note also that the frequency in the rest frame of the compact object is conserved at each graviton-worldline vertex, owing to the integral over $d\tau$, but momentum is not conserved explicitly since we are treating the worldline as a heavy source against which gravitons are scattered. Note, however, that this is accounted for by the recoil term in the action in Eq.~\eqref{eq:action_expanded}.
\end{widetext}

The tidal response function, which describes the dynamics of the induced quadrupole is given by $F(\omega)$, and enters the scattering problem via the propagator for the quadrupole moment as shown in Fig.~\ref{fig:Frules}. It is assumed to be invariant under $\omega\rightarrow -\omega$ and purely real. Note that it appears alongside the tensor
\begin{equation}
    \mathcal{P}^{\mu\nu,\rho\sigma} = \frac{1}{2}(\mathscr{P}^{\mu\rho}\mathscr{P}^{\nu\sigma}+\mathscr{P}^{\mu\sigma}\mathscr{P}^{\nu \rho} - \frac{2}{3} \mathscr{P}^{\mu\nu}\mathscr{P}^{\rho\sigma}).
\end{equation}
This ensures that the $\langle QQ \rangle $ propagator is symmetric, traceless and orthogonal to the 4-velocity $u^\mu$, so that the dynamics of the tidal quadrupole moment preserve the constraints $Q^{\mu\nu}u_\nu = 0$, $Q^\mu{}_\mu=0$, at leading order in the metric perturbation. This way of defining the tidal response is completely equivalent to an explicit ansatz relating the frequency modes of the tidal fields $E^{\mu\nu}$ and $Q^{\mu\nu}$~\cite{Goldberger:2005cd,Goldberger:2020fot,Saketh:2023bul}.

\begin{figure}[H]
    \centering
    \includegraphics[width=0.5\linewidth]{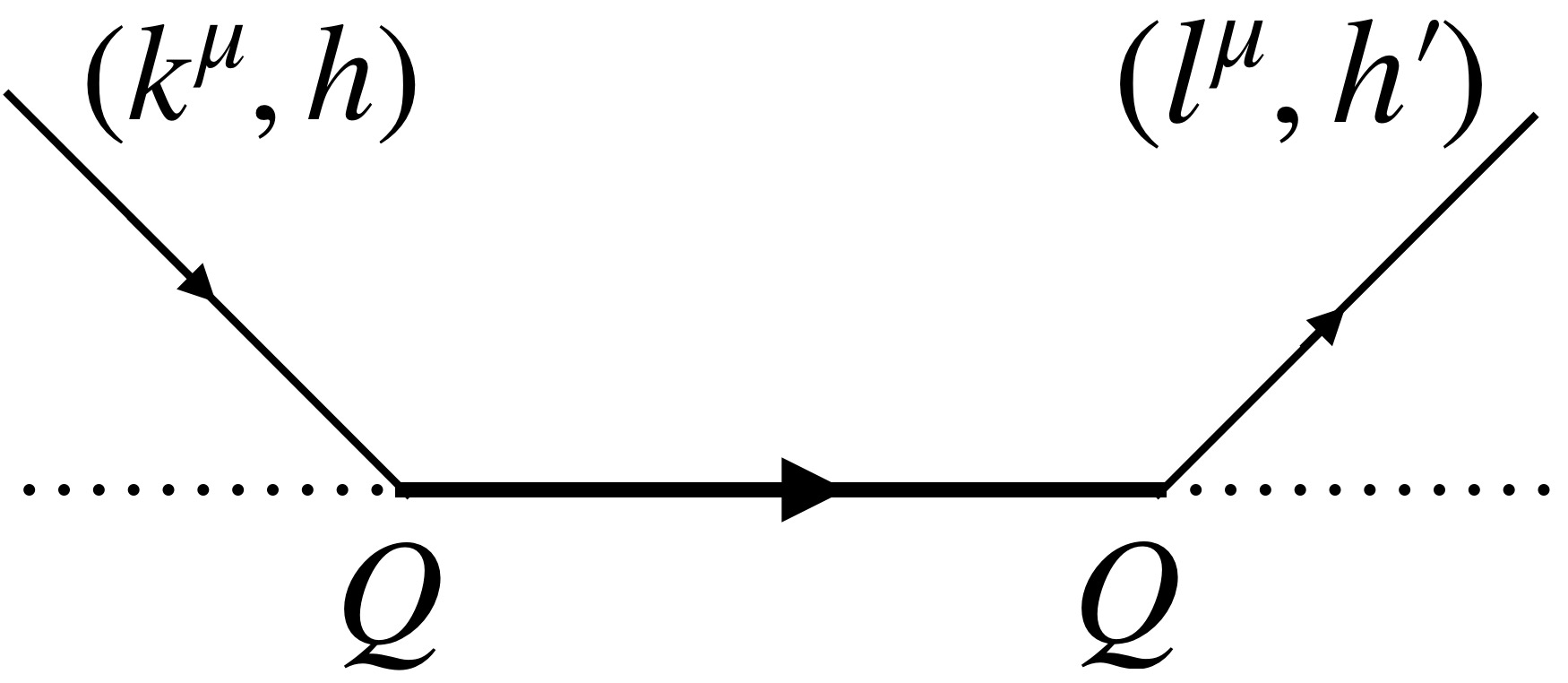}
    \caption{Tree level, tidal contribution to Raman scattering}
    \label{fig:treeQQ}
\end{figure}

We can now write down the leading order contribution to the Raman scattering process due to tidal effects, shown in Fig.~\ref{fig:treeQQ}, as 
\begin{multline}
%\begin{split}
         i\mc{M}_{\omega;\vec{k},h\rightarrow \vec{l},h'}|_{\rm tree~level} = -\frac{1}{m_{pl}^2} u^{\mu}u^{\sigma}k_{[\mu} \epsilon^{h'}_{\nu][\rho}(k)k_{\sigma]} \\
         \times i R^5 F(\omega) \mc{P}^{\nu \rho,\beta \gamma} \times u^{\alpha}u^{\delta} l_{[\alpha}\bar{\epsilon}^h_{\beta][\gamma}(l)l_{\delta]}
        \\ = \frac{-iR(R\omega)^4F(\omega)}{16 m_{pl}^2}\epsilon_{ij}^{h}(\vec{k})\bar{\epsilon}^{ij}_{h'}(\vec{l}) %,~\omega = -k\cdot u,~-l\cdot u,
%\end{split}
\label{eq:MtreeC}
\end{multline}
where in the second relation we have chosen the polarization tensors in  traceless-transverse gauge, i.e., $\epsilon^{0\mu}=0$,~$\epsilon^{\mu}{}_{\mu}=0$. We have also defined $\omega=k^0=-k\cdot u=l^0=-l\cdot u$. Note that the covariant notation makes gauge invariance of the amplitude manifest in the first line in Eq.~\eqref{eq:MtreeC}, since it is invariant under $\epsilon_h^{\mu\nu}(k)\rightarrow \epsilon_h^{\mu\nu}(k)+2\zeta^{(\mu} k^{\nu)}$ (and similarly for $\epsilon^{\mu\nu}_h(l)$) thereby satisfying the gravitational Ward identity (or equivalently, invariance under linearized diffeomorphisms/coordinate transformations).

\begin{figure}[H]
    \centering    \includegraphics[width=0.8 \linewidth]{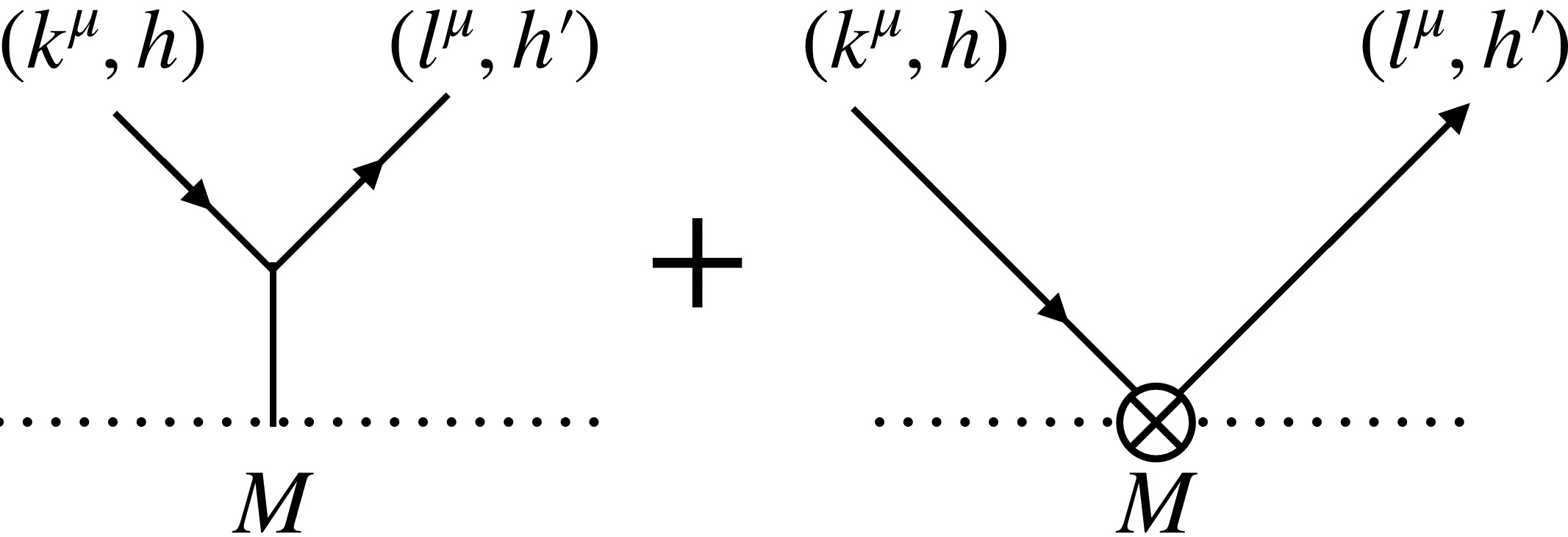}
    \caption{Tree level, non-tidal contributions to Raman scattering. The $\otimes$ vertex is due to the recoil term in Eq.~\eqref{eq:action_expanded}, and vanishes in traceless transverse gauge. However, more generally, this contribution is essential to preserve gauge invariance.}
    \label{fig:treeTT}
\end{figure}

 However, this is far from the only, or even the most relevant diagram for the scattering processes. For instance, there are several diagrams, such as those shown in Fig.~\ref{fig:treeTT}, which do not involve any tidal effects. We will shortly show how to systematically include or discard them as needed.

 However, let us first understand the manner in which the contribution indicated in Fig.~\ref{fig:treeQQ} (or other such EFT diagrams) can be matched with the scattering phase in Eq.~\eqref{eq:SPT_phase}. To facilitate this comparison, we need to switch to a partial-wave basis. The Feynman diagrams naturally provide matrix elements of the S-matrix between fixed momentum and helicity states, i.e., 
 \begin{equation}
     i\langle \vec{l},h'| T |\vec{k},h\rangle = i\mc{M}(\omega;\vec{k},h\rightarrow \vec{l},h')2\pi \delta(|\vec{k}|-|\vec{l}|).
 \end{equation}
This can be converted to the $\ell,m,h$ basis using the relations~\cite{Saketh:2023bul}
\begin{equation}
%    \begin{split}
        |\vec{k},h\rangle =  \sum_{\ell=2}^\infty \sum_{m=-\ell}^{\ell} 2\pi \sqrt{\frac{2\ell+1}{2\pi |\vec{k}|}} D^{\ell}_{m h}(\hat{k},0) |\omega=|\vec{k}| ,\ell,m, h\rangle, 
\end{equation}
and
\begin{equation}
        \epsilon^{h=\pm 2}_{ij} = \sum_{m=-2}^{m=2} D_{mh}^{\ell=2} \langle i,j| \ell=2,m\rangle,
%    \end{split}
\end{equation}
where $\langle i,j | \ell=2,m \rangle$ are the expansion coefficients of the usual spherical harmonics in the symmetric tracefree (STF) tensor basis. Thus, we have 
\begin{equation}
    i\langle\omega'\ell m h'| T | \omega\ell mh\rangle = i \mc{M}_{\omega,\ell;h\rightarrow h'} 2\pi \delta(\omega-\omega'),
\end{equation}
where
\begin{equation}
    i\mc{M}_{\omega,\ell;h\rightarrow h'} = \frac{-i(R\omega)^5F(\omega)}{160 m_{pl}^2 \pi},
\end{equation}
at tree-level. Note that the above result does not depend on helicity. Finally, we can express the result in the parity basis, which is the diagonal partial-wave basis, using 
\begin{equation}
\begin{split}
        |\omega\ell m P\rangle & = \frac{|\omega\ell m h=+2\rangle  \pm (-1)^\ell |\omega\ell m h=-2\rangle}{\sqrt{2}}.
\end{split}
\end{equation}
In the parity basis, we then have
\begin{equation}
     i\langle \omega' \ell m P=+ |  T | \omega \ell m P=+\rangle = i \mc{M}_{P\omega\ell} 2\pi \delta(\omega-\omega'), \label{eq:treeQQ} 
\end{equation}
with
\begin{equation}
   i\mc{M}_{P\omega\ell}= \frac{-i(R\omega)^5 F(\omega)}{80 m_{pl}^2 \pi},
\end{equation}
at tree-level. Finally, we use the operator relation, $S=I+iT$, to define the scattering phase in EFT as 
\begin{multline}
%\nonumber
        2\pi \delta(\omega-\omega')e^{2 i\delta_{\ell \omega}^{\rm EFT}} = i\langle \omega' \ell m P=+ |  S | \omega \ell m, P=+\rangle \\ = 2\pi \delta(\omega-\omega')+ i \langle \omega' \ell m P 
        =+ |  T | \omega \ell m P=+\rangle.  
\end{multline}
Noting that we restrict ourselves to positive parities, this can be rewritten as 
\begin{equation}
   e^{2i\delta_{P=+,\omega}^{\ell m}} =1+ i \mc{M}(\omega,\ell,P).  
\end{equation}
It is also convenient to define a new operator $\Delta$, as shown in Ref.~\cite{Ivanov:2026icp}:
\begin{equation}
%\begin{split}
        e^{2i\Delta} = S = I + i T, 
        \label{eq:amp2phase}
\end{equation}
with
\begin{equation}
    2\pi \delta(\omega-\omega') \delta_{\ell\omega}^{\rm EFT}= \langle \omega'\ell m |\Delta |\omega \ell m\rangle.
            \label{eq:amp2phase2}
\end{equation}
In order for the EFT to be a faithful description of the stellar response, we expect $\delta^{\rm EFT}$ to be identical to the phase obtained from perturbation theory in Eq.~\eqref{eq:SPT_phase}.

This concludes the discussion of how the amplitudes can be related to the scattering phase. However, there is a major inefficiency associated with the calculation of the right-hand side of equation Eq.~\eqref{eq:amp2phase}. As mentioned before, $T$ does not just receive contributions from diagrams containing tidal effects, i.e., the quadrupole moment. It also involves diagrams representing the process of waves scattering off the Schwarzschild background of the unperturbed star, e.g., the first diagram in Fig.~\ref{fig:treeTT}. This greatly increases the required effort. In fact, one has to compute very high-loop diagrams corresponding to non-tidal (also referred to as far-zone) scattering processes in order to successfully use Eq.~\eqref{eq:amp2phase} to fix the tidal Love number~\cite{Saketh:2023bul,Ivanov:2022qqt}. Efforts to compute the required high-loop, far-zone diagrams are still underway~\cite{Ivanov:2026icp, Chang:2026eti,Caron-Huot:2025tlq, Ivanov:2024sds}. 

For nonspinning neutron stars, this issue can be avoided by replacing $\Delta_{\rm NS}$ with a different Hermitian operator that (i) isolates the dynamical tidal response of the star and (ii) possesses matrix elements that can be directly related to the tidal phase shift $\delta^{\rm tidal}_{\ell\omega}$ appearing in Eq.~\eqref{eq:SPT_tidal_phase}. We will argue below that these requirements are satisfied by defining
\begin{equation}
    \tilde{\Delta}_{\rm tidal}^{\rm NS} = \Delta_{\rm NS} - \Re\Delta_{\rm BH}.
\end{equation}
Then, through the relation of the scattering phase $\delta$ to the operator $\Delta$ in Eq.~\eqref{eq:amp2phase2}, the diagonal matrix elements of $\tilde{\Delta}_{\rm tidal}^{\rm NS}$ can be  matched to $\delta^{\rm tidal}_{\ell \omega}$ using Eq.~\eqref{eq:amp2phase}. Further, $\Re\Delta_{\rm BH}$ should be understood as the corresponding operator for an EFT which has the same conservative dynamical tidal response as a black hole, but no dissipation (e.g., absorption associated with the event horizon). 

However, it is still not clear how we expect to compute $\tilde{\Delta}_{\rm tidal}^{\rm NS} $ in the EFT without working out the high-order, far-zone, loop diagrams. Also, as we do not yet know the dynamical tidal response of a black hole, how do we use the matching to ultimately fix $F(\omega)$ which enters $\Delta_{\rm NS}$? We will discuss these two questions in the following.

We start by defining (labeling) $\exp(\Re\Delta_{\rm BH})=1+i T_{\rm bh}$, and $\exp\Delta_{\rm NS}=1+ i T_{\rm NS}$, in accordance with Eq.~\eqref{eq:amp2phase}. We can then write
\begin{equation}
    \begin{split}
        &e^{2i(\Delta_{\rm NS}-\Re\Delta_{\rm BH})} = (1+i T_{\rm NS})(1-i T_{\rm bh}^\dagger)\\&=1+i(T_{\rm NS}-T_{\rm bh}^\dagger)+T_{\rm NS}T_{\rm bh}^\dagger. 
    \end{split}
    \label{eq:tidal_op1}
\end{equation}
At this point, we mathematically implement the approximation 
\begin{equation}
\begin{split}
        &T_{\rm NS} \rightarrow T_{\rm pp}+T_{\rm tidal},
        \\& T_{\rm bh}\rightarrow T_{\rm pp},
\end{split}
\label{eq:prescription}
\end{equation}
where $T_{\rm pp}$ is the non-trivial scattering operator for an EFT without tidal effects, i.e., no quadrupolar degree of freedom.

The idea is to approximate the black hole as a compact object with vanishing dynamical tidal response. Two further questions immediately arise: (i) To what extent can a black hole's dynamical tidal response be ignored?, and (ii) Is this manner of implementing the approximation consistent? It is straightforward to answer the first question. From Eq.~\eqref{eq:treeQQ}, it is clear that quadrupolar tidal effects first contribute at order $(M\omega)^5$ for black holes. However, owing to the vanishing static tidal response~\cite{Binnington:2009bb,Damour:2009vw}, the leading quadrupolar tidal effects are dissipative and actually enter at  order $(M\omega)^6$. Given that $T_{\rm bh}$ only encodes conservative contributions to the scattering process (since it is defined using $\Re(\Delta_{\rm BH})$), it does not encode dissipation and thus tidal effects contribute only starting from $(M\omega)^7$. Comparing this to the leading order scaling for a neutron star in Eq.~\eqref{eq:amp2phase}, it is clear that the dynamical tidal response of a black hole is suppressed by $(M/R)^5(M\omega)^2$, which  can be neglected (at least for our present purposes)\footnote{In the PN framework, this represents an error at order 3PN if we identify $\omega$ with approximately the orbital frequency, along with a suppression due to the large power of $M/R<1$.}. One may object,  suggesting that a different parameterization of the $\langle QQ\rangle $ propagator in Fig.~\ref{fig:Frules} could have led to a prefactor of $(M\omega)^5$ in Eq.~\eqref{eq:treeQQ}, in which case the suppression is just $(M\omega)^2$. This is true, but the parameterization in Fig.~\ref{fig:Frules} was chosen in accordance with known behaviour of the static tidal deformability of neutron stars, which approximately scales as the fifth power of compactness ($C=M/R)$~\cite{YagiYunes2017,Maselli2013}.

The second question, as to whether Eq.~\eqref{eq:prescription} is the mathematically correct way to implement the argument, is more subtle. The suggested replacement does not even hold approximately at all orders in $M\omega$. This is since $T_{\rm bh}$ is a finite and well-defined operator, whereas $T_{\rm pp}$ has divergences owing to far-zone, multi-loop diagrams starting at order $(M\omega)^7$~\cite{Saketh:2023bul}. These divergences need to be regulated, and then cancelled by counter terms that are essentially tidal in nature, and thus the EFT of a point particle without tides is not meaningful at high orders in $(M\omega)$. To alleviate this issue, we can instead rewrite Eq.~\eqref{eq:prescription} using manifestly finite operators as
\begin{equation}
\begin{split}
        &T_{\rm NS} \rightarrow \tilde{T}_{\rm pp}+\tilde{T}_{\rm tidal},
        \\& T_{\rm bh}\rightarrow \tilde{T}_{\rm pp},
\end{split}
\label{eq:regprescription}
\end{equation}
where $\tilde{T}_{\rm pp}$ and $\tilde{T}_{\rm tidal}$ are obtained by subtracting the divergent pieces\footnote{Divergences in $T_{\rm pp}$ are due to multi-loop diagrams starting at $(M\omega)^7$, whereas the divergences in $T_{\rm tidal}$ are due to counter terms included to cancel out the divergences in $T_{\rm pp}$. There are also loop divergences in $T_{\rm tidal}$ which are discussed further below. They are not of concern here.} in $T_{\rm pp}$ and $T_{\rm tidal}$  via counter terms in the latter (possibly leaving behind some finite pieces scaling as $(M\omega)^7$ or higher with $\mc{O}(1)$ coefficients). In the first replacement, we can in fact choose these counter terms so that they cancel out in the sum (leaving behind a logarithmic running dependence on the gravitational-wave frequency in the amplitudes), and the result is therefore exact. In the second replacement, however, since we do not know the dynamical tidal response of black holes, we are introducing errors  of order $(M\omega)^7$ and higher, but unless the coefficients accompanying these terms are unexpectedly large, these errors are expected to be small.

Accepting these arguments as valid, and using the unitarity of $1+i\tilde{T}_{\rm pp}$, we can rewrite Eq.~\eqref{eq:tidal_op1} as
\begin{equation}
%   \begin{split}
        e^{2i(\Delta_{\rm NS}-\Re\Delta_{\rm BH})}=e^{2i \Delta_{\rm tidal}} =1+i \tilde{T}_{\rm tidal}+\tilde{T}_{\rm 
        tidal}\tilde{T}_{\rm pp}^\dagger. 
%    \end{split}
    \label{eq:tidal_op_t}
\end{equation}
The right-hand side of this equation contains the tidal contribution at leading order. Although the third term also involves the point-particle amplitude, it always appears multiplied by the tidal contribution. Consequently, a consistent calculation no longer requires computing the point-particle amplitude to high loop orders.

\begin{figure}[h!]
        \centering
        \includegraphics[width=0.8\linewidth]{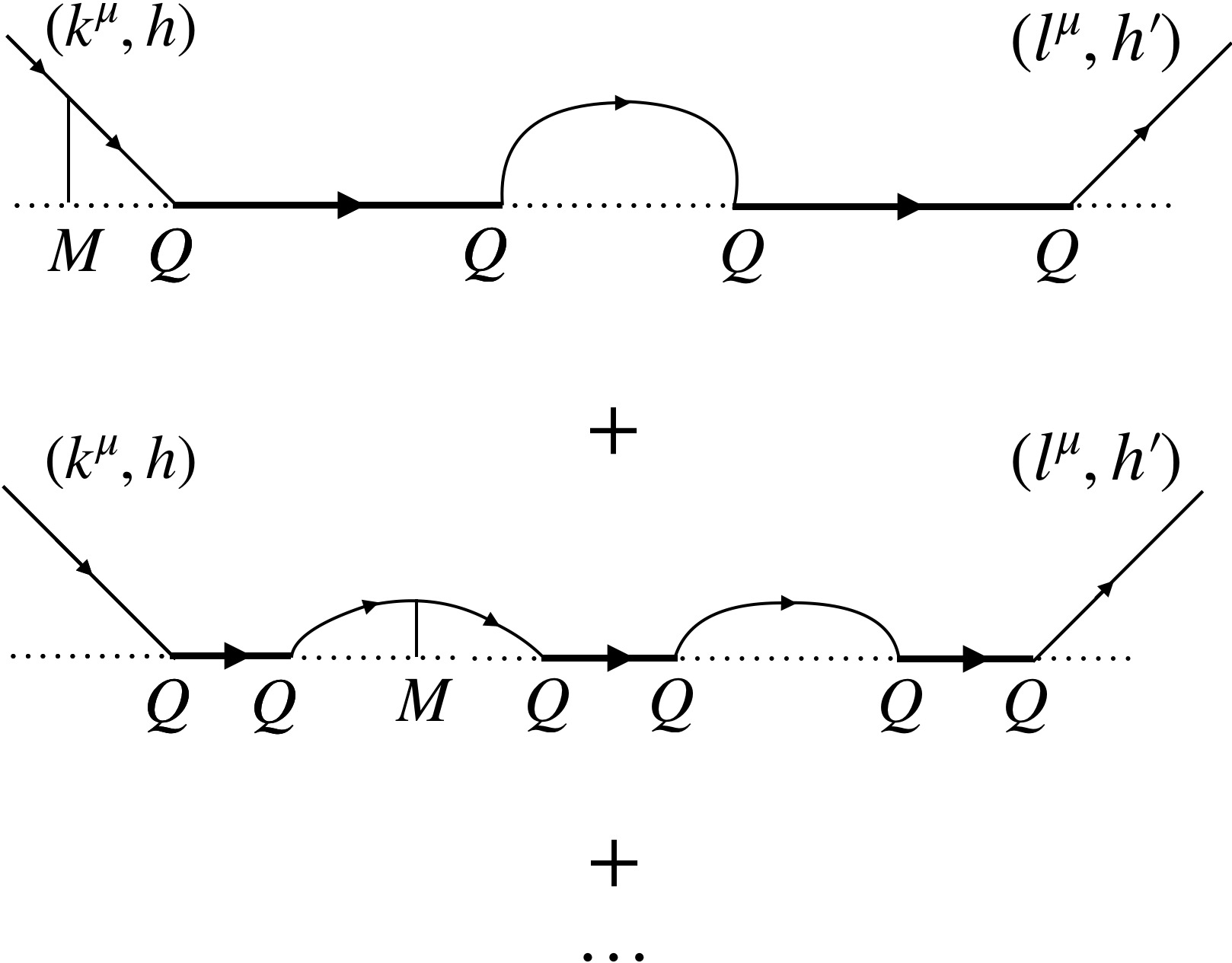}
        \caption{Additional contributions to $T_{\rm tid,(1)}$ involving one mass insertion.}
        \label{fig:treeMQQchains}
    \end{figure}

Finally, it is convenient to expand the operators $T_{\rm BH}^\dagger$ in powers of $M$ as 
\begin{align}
&\tilde T_{\rm pp}
    = M T_{(1)} + M^2 T_{(2)} + \mathcal{O}(M^3),\\& \tilde T_{\rm tidal}
    = T_{\rm tid,(0)}
     + M T_{\rm tid,(1)}
     + \mathcal{O}(M^2),
\end{align}
Substituting these expansions gives
\begin{multline}
e^{2 i \Delta_{\rm tidal}}
=1+iT_{\rm tid,(0)}
\\
+M\!\left(iT_{\rm tid,(1)} 
+T_{\rm tid,(0)}T_{(1)}^\dagger\right)
  +\mathcal O(M^2).
  \label{eq:expT}
\end{multline}
This further illustrates the advantage of the subtraction scheme. The $\mathcal{O}(M)$ term above determines the $\mathcal{O}(M\omega)$ correction to the tidal response. Naively, obtaining this correction would require computing the point-particle amplitude through $\mathcal{O}((M\omega)^6)$, corresponding to five-loop diagrams. The subtraction scheme instead reduces the required calculation to one loop (two-loops for certain diagrams, see Fig.~\ref{fig:treeMQQchains}). The same simplification extends to higher orders in $M$.

Let us now evaluate the contributions at each order in $M$ appearing on the right-hand side of Eq.~\eqref{eq:expT}. In this work, we restrict our attention to the leading and next-to-leading terms, corresponding to orders $M^0$ and $M^1$. In view of this restriction, the earlier discussion of far-zone divergences is not strictly necessary for the present analysis. Nevertheless, it lays prepares the ground for future extensions to higher orders.

\begin{figure}[h]
    \centering
    \includegraphics[width=0.8\linewidth]{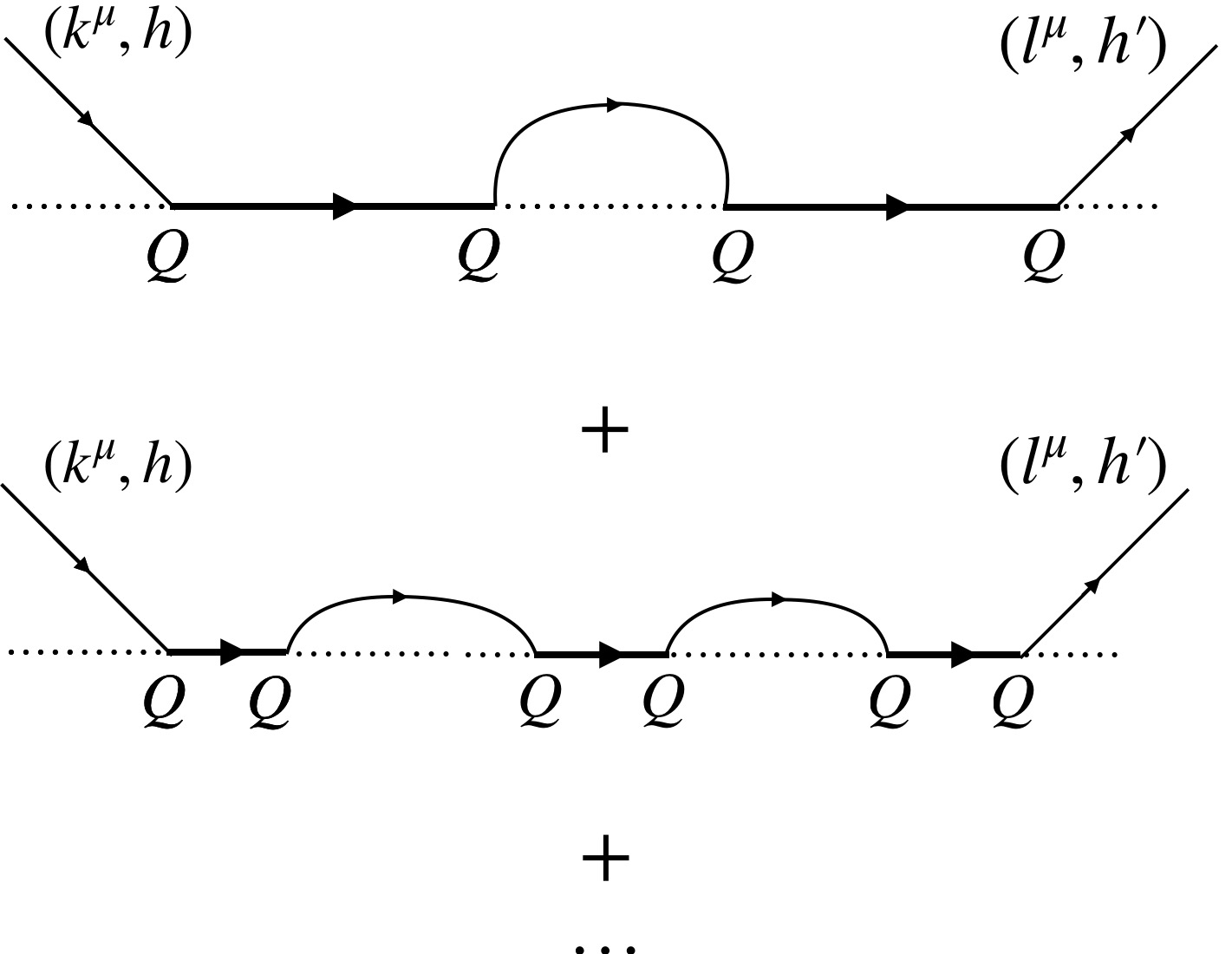}
    \caption{Additional tidal contributions to the scattering process that do not involve mass insertions such as those in Fig.~\ref{fig:treeMQQ}. These diagrams effectively resum the result of Fig.~\ref{fig:treeQQ}, and are required to account for backreaction due to gravitational radiation.}
    \label{fig:treeQQsummed}
\end{figure}

    The tree-level contribution to $T_{\rm tid,(0)}$ is simply given by the diagram in Fig~\ref{fig:treeQQ}. However, we can get other contributions that are independent of $M$ by chaining together such diagrams as shown in Fig.~\ref{fig:treeQQsummed}. 
Summing over all diagrams indicated in Fig.~\ref{fig:treeQQsummed} yields the net tidal amplitude at $M^0$. Referring to the contribution of Fig~\ref{fig:treeQQ} as $T^{\rm tree}_{\rm tid,(0)}$, we obtain\footnote{The diagrams are formally divergent, but can be regulated through dimensional regularization to obtain the result presented in the text. In other schemes, they can lead to additional     contributions which can be absorbed into a redefinition of $F(\omega)$ to once again obtain Eq.~\eqref{eq:treesummed}.}

\begin{equation}
    T_{\rm tid,(0)} = \frac{T^{\rm tree}_{\rm tid,(0)}}{1-i \frac{T^{\rm tree}_{\rm tid,(0)}}{2}}.
    \label{eq:treesummed}
\end{equation}

As a sanity check, we can use the unitarity of $\exp(i\Delta_{\rm tidal}$) at leading order in $M$ to obtain the constraint $2 \Im(T_{\rm tid,(0)}) = T_{\rm tid,(0)}T_{\rm tid,(0)}^{\dagger},$ from Eq.~\eqref{eq:expT}. It is easy to check that Eq.~\eqref{eq:treesummed} satisfies this requirement, using the fact that $T^{\rm tree}_{\rm tid,(0)}$ is Hermitian\footnote{This follows from assuming that there is no viscosity, meaning $F(\omega)$ can only be imaginary at resonant poles. The latter is insufficient by itself if there is a continuum of resonant modes (i.e., gapless degrees of freedom), and thus we further assume that the spectrum is discrete in the relevant regime of $\omega$ in this work.}. 

\begin{figure}[h!]
        \centering
        \includegraphics[width=0.9\linewidth]{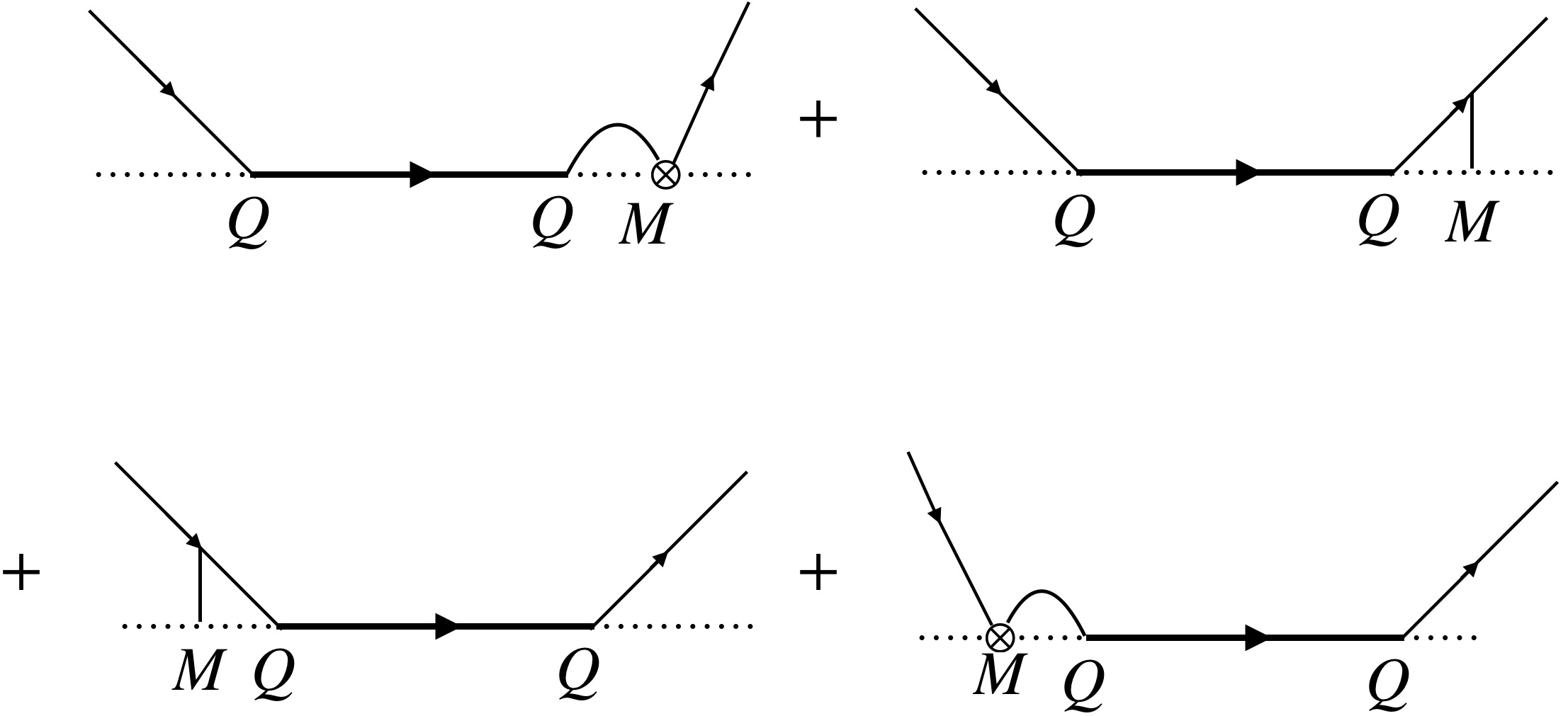}
    \caption{The simplest diagrams at linear order in $M$. The loop leads to an infrared divergence as well, which does not show up in any observables. The labels for incident and outgoing momenta have been suppressed to avoid cluttering. }
        \label{fig:treeMQQ}
    \end{figure}

    Moving on to linear order in $M$ on the right-hand side of Eq.~\eqref{eq:expT}, the first contribution to $T_{\rm tid,(1)}$ is given in Fig.~\ref{fig:treeMQQ}. This diagram requires the three-graviton vertex, but fortunately, it can be evaluated using the ingredients already available in Ref.~\cite{Goldberger:2009qd}.

     As before, this diagram can be chained together with multiple copies of the diagram in Fig.~\ref{fig:treeQQ} to create many more contributions to $T_{\rm tid,(1)}$ such as those shown in Fig.~\ref{fig:treeMQQchains}. 

Strictly speaking, we should use the Feynman rule for the 3-graviton vertex to compute such chains to obtain a summed version. However, we leave a proper analysis of this for the future and instead ``guess'' a form that is consistent with the leading order piece and satisfies the unitarity constraints. The required result  takes the form
\begin{equation}
    T_{\rm tid,(1)}  = \frac{\Im(T^{\rm simple}_{\rm tid,(1)})}{1-i \frac{T^{\rm tree}_{\rm tid,(0)}}{2}} + \frac{\Re(T^{\rm simple}_{\rm tid,(1)})}{\left(1-i\frac{T^{\rm tree}_{\rm tid,(0)}}{2}\right)^2}
    \label{eq:treeMsummed}
\end{equation}
where we refer to the contribution from  Fig.~\ref{fig:treeMQQ} as\footnote{$T_{\rm tid,(1)}$ receives contributions from loop diagrams, and thus we cannot refer to the simplest diagram as a tree diagram.} $T^{\rm simple}_{\rm tid,(1)}$. Now, defining 
\begin{equation}
    G_{(1)} = T_{\rm tid,(1)} - i T_{\rm tid,(0)} T_{(1)}^\dagger,
\end{equation}
the unitarity of $\exp(i\Delta_{\rm tidal}$) at linear order in $M$ yields
\begin{equation}
    2 \Im G_{(1)} = T_{\rm tid,(0)}G_{(1)}^\dagger + G_{(1)}T_{\rm tid,(0)}^\dagger.
\end{equation}
This is satisfied by Eq.$~\eqref{eq:treeMsummed}$ provided $\Im(T_{\rm tid,(1)}^{\rm simple}) = T^{\rm tree}_{\rm tid,(0)}T_1$, which in turn follows trivially from taking the imaginary part of the sum of diagrams in Fig.~\ref{fig:treeMQQ} with the help of the cutting rules. We have also used $T_{(1)}^\dagger = T_{(1)}$ to obtain this relation, which follows simply from the fact that $T^{(1)}$ is the contribution due to the diagrams in Fig.~\ref{fig:treeTT}, which have no propagators that can be cut. 

At this point we can rewrite
\begin{equation}
    G_{(1)} = T_{\rm tid,(1)} - i T_{\rm tid,(0)} T_{(1)}^\dagger = \frac{\Re(T_{\rm tid,(1)}^{\rm simple})}{\left(1-i \frac{T^{\rm tree}_{\rm tid,(0)}}{2}\right)^2}.
\end{equation}

Finally, defining a frequency operator $\hat{\omega}$ as $\hat{\omega}|\vec{k},h\rangle = |\vec{k}| |\vec{k},h\rangle$, and using the results of Ref.~\cite{Goldberger:2009qd}, we can write the operator relation
\begin{equation}
        \Re(T_{\rm tid,(1)}^{\rm simple}) = T^{\rm tree}_{\rm tid,(0)} 2\pi \hat{\omega}.
\end{equation}

The analysis can be extended straightforwardly (albeit with increasing tediousness) to higher orders in $M$. However, here we restrict our attention to linear order in $M$. Combining the results, we obtain
\begin{equation}
    e^{2i\Delta_{\rm tidal}} = 1+i\frac{T^{\rm tree}_{\rm tid,(0)}}{1-i \frac{T^{\rm tree}_{\rm tid,(0)}}{2}}\left( 1+\frac{2\pi M\hat{\omega}}{1-i \frac{T^{\rm tree}_{\rm tid,(0)}}{2}}\right)
    \label{eq:presummed}
\end{equation}
which is unitary to $\mc{O}(M^2)$. This can be improved by simply resumming Eq.~\eqref{eq:presummed} by adding terms $\mc{O}(M^2)$ and higher as
\begin{equation}
    e^{2i\Delta_{\rm tidal}} = \frac{1+i \frac{T^{\rm tree}_{\rm tid,(0)}}{2}(1+2 \pi M \hat{\omega})}{1-i \frac{T^{\rm tree}_{\rm tid,(0)}}{2}(1+2 \pi M \hat{\omega})},
    \label{eq:resummed}
\end{equation}
which is unitary to all orders in $M$. Now, using
    \begin{equation}
    \delta_{\ell\omega}^{\rm EFT,\rm tidal} = \frac{\langle \omega,\ell,m, P| \Delta_{\rm tidal}|\omega_1, \ell,m ,P\rangle}{2\pi \delta(\omega-\omega_1)},
    \end{equation}
  and the result for the tree-level tidal phase from Eq.~\eqref{eq:treeQQ}, we obtain the scattering phase  as
\begin{equation}e^{2i\delta_{\ell\omega}^{\rm EFT,\rm tidal}} = \frac{1-i\frac{(R\omega)^5 F(\omega)}{160 m_{pl}^2 \pi}(1+2\pi M|\omega|)}{1+i\frac{(R\omega)^5 F(\omega)}{160 m_{pl}^2 \pi}(1+2\pi M|\omega|)}
\label{eq:EFTphasexp}
\end{equation}
This is the result we need to match with the scattering phase from perturbation theory, given earlier in Eq~\eqref{eq:SPT_tidal_phase}, in order to determine the tidal response function, $F(\omega)$.

\section{Results}
\label{secIV:RES}

After studying the Raman scattering process in the effective theory and exploring how the problem connects to  neutron-star perturbation theory, we arrived at two expressions for the tidal scattering phase; Eqs.~~\eqref{eq:SPT_tidal_phase} and \eqref{eq:EFTphasexp}. Matching these, we obtain the tidal response function as
\begin{equation}
    F(\omega) = -\frac{160 m_{pl}^2 \pi}{(R\omega)^5(1+2 \pi M |\omega|)}\tan\delta_{\ell\omega}^{\rm tidal}.
\end{equation}
It is convenient to relate this result to the usual definition of the dimensionless tidal response as $k_2(\omega)=-(3/4)F(\omega)$. This leads to
\begin{multline}
        k_2^{\rm EFT}(\omega) =  \frac{120 m_\mathrm{pl}^2 \pi}{(R\omega)^5(1+2 \pi M |\omega|)}\tan\delta_{\ell\omega}^{\rm tidal}
        \\ = \frac{15 }{4 (R\omega)^5(1+2 \pi M |\omega|)}\tan\delta_{\ell\omega}^{\rm tidal},
        \label{eq:dynamical_tides}
\end{multline}
where $e^{2i\delta_{\ell \omega}^{\rm tidal}}$ is obtained from the analytical MST solutions using Eq.~\eqref{eq:SPT_phase}. 

We naturally want to quantify the tidal response we have derived and compare with previous results. This is, however, problematic as no appropriately validated results in this direction exist. A rigorous assessment  would require computing the corresponding gravitational waveforms (or the corresponding phase evolution) to facilitate a direct comparison. Given that we would then need to connect to the orbital evolution, this would require a fair amount of extra work which we leave for the future. It is also relevant to note there are currently no validated models that account for the presence of low-frequency g-modes in the tidal response. State-of-the-art waveform models~\cite{Hinderer2016EffectsON,NRttidalv3} only account for the dynamical tide of the fundamental neutron-star mode, which is not expected to pass through resonance before binary merger.  Additionally, typical waveform models involve free parameters that are tuned to match numerical simulations.  Hence, it is unclear if such a comparison would actually help distinguish between expressions for the tidal response.
What we actually need is progress on alternatives to the problem, such as the relativistic mode-sum approach proposed in Ref.~\cite{HegadeKR:2026kku}, and then to come up with observables that can be consistently compared across different methods. Once completed, the mode-sum approach would allow an apples-to-apples comparison, which would be very welcome.

As an approximate diagnostic, we may  contrast our results with earlier expressions for the dynamical tidal response, such as those from Ref.~\cite{Andersson:2025iyd}. Admittedly, this comparison assumes that the overall strategy to infer the tidal response through matching makes sense. Even this comparison is not entirely straightforward, since the tidal responses are defined somewhat differently. However, the comparison  highlights the qualitative features; particularly, the behavior near resonances, and we can make meaningful statements regarding the improvements brought by the present analysis. As an example of this we can explore the low-frequency limit in which the response is characterized by a single static Love number, common to all approaches up to normalization conventions. In fact, a crucial benchmark for any dynamical tide model is how accurately the low-frequency behavior reproduces the  static limit. As we will demonstrate, our model performs very well in that respect.

\subsection{Tidal response near quasinormal modes}

We start by considering the dynamical tidal response $k_2(\omega)$ obtained from Eq.~\eqref{eq:dynamical_tides} near a few resonant modes for a specific neutron-star model. For this example, we focus on the BSk22 equation of state from the Brussels-Montreal collaboration~\cite{2013PhRvC..88f1302G,2018MNRAS.481.2994P,BSkGR}. This is a convenient choice because we can then make direct comparisons with the results from~\cite{Andersson:2025iyd}. We will not provide results for other models here, because we only aim for a proof-of-principle demonstration at this stage.

\begin{figure}[h]
\includegraphics[width=\linewidth]{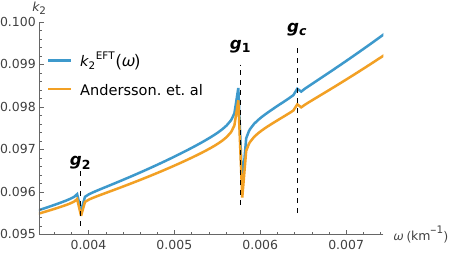}

    \caption{Illustrating the  dynamical tidal response near resonant frequencies; in this case the highest frequency g-modes for the chosen BSk model. The tidal response obtained from Eq.~\eqref{eq:dynamical_tides} is labeled ``$k_2^{\rm EFT}(\omega)$'', while the result from Ref.~\cite{Andersson:2025iyd} is labeled ``Andersson. et. al''. Given that the two sets of results are very close, the considerably more advanced formalism we have brought appears to have only a modest effect on the tidal response near the g-mode resonances. }
    \label{fig:four_panel_near_modes}
\end{figure}

\begin{figure}[h!]
    \centering
    \includegraphics[width=\linewidth]{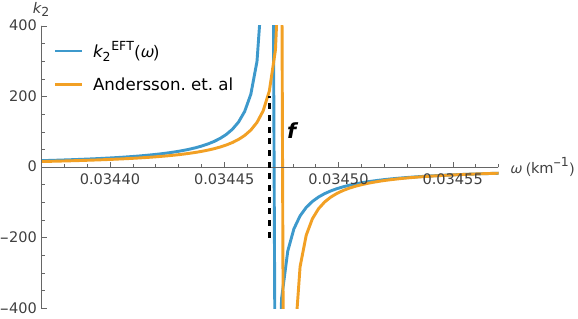}
    \caption{A zoomed in plot providing a detailed comparison near the dominant f-mode. Both tidal responses roughly show resonant behaviour close to the QNM frequency. The new result seems to align slightly more closely with the f-mode, but the shift is quite small.}
    \label{fig:near_f_zoom}
\end{figure}

Our first model is a neutron star with mass $M=1.4~M_{\odot}$ and radius $R=13.04$~km. A few of relevant  quasinormal modes (QNMs) for this model are $\omega_{g_{\rm c}}=0.00643~\mr{km}^{-1}$, $\omega_{g_1} = 0.00577~\mr{km}^{-1}$, $\omega_{g_2}=0.0039~\mr{km}^{-1}$, and $\omega_f=0.0345~\mr{km}^{-1}$. In Fig.~\ref{fig:four_panel_near_modes}, we illustrate the tidal response near a couple of the g-modes. Meanwhile, Fig.~\ref{fig:near_f_zoom} shows the tidal response near the f-mode.

It is evident that Eq.~\eqref{eq:dynamical_tides} is sensitive to resonant features and the results in Fig.~\ref{fig:four_panel_near_modes} and \ref{fig:near_f_zoom}  closely resemble the results from~\cite{Andersson:2025iyd}. This is an important consistency check, since one of the primary applications of the dynamical tidal response is to capture the nontrivial dynamics that arise when the orbital frequency enters resonance with a stellar oscillation frequency. As expected, we find that the strongest resonant behavior occurs near the $f$-mode, see Fig.~\ref{fig:near_f_zoom}. This demonstrates why the f-mode is the only mode currently incorporated into waveform models. Its dominance is evident in all dynamical tide calculations. Overall, the results indicate that the tidal response obtained from our new model 
differs only slightly from previous results in Ref.~\cite{Andersson:2025iyd}. If anything, the results show that the analysis in~\cite{Andersson:2025iyd} was already rather good (at least for less massive neutron stars).

\begin{figure}[h!]
    \centering
    \includegraphics[width=\linewidth]{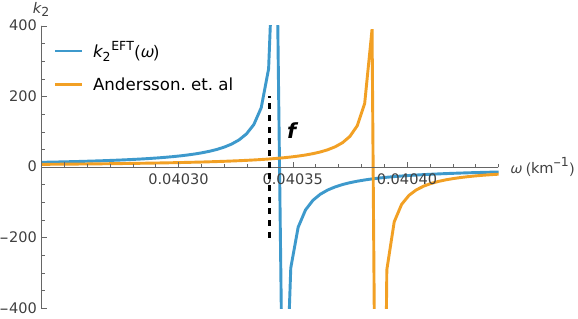}
    \caption{A zoomed in plot providing a detailed comparison near the dominant f-mode for $M=1.98M\odot$, and $R=12.59$~km. Note that $k_2(\omega)$ shows resonance behaviour very close to the expected  f-mode frequency (indicated by the dashed, vertical red line) whereas the earlier result in Ref.~\cite{Andersson:2025iyd} shows a slight offset towards higher frequencies.}
    \label{fig:near_f_zoom_m198}
\end{figure}

While the f-mode frequencies (the location of the tidal resonance) are very close, Fig.~\ref{fig:near_f_zoom} suggests that the new model moves slightly closer to the known value of the mode frequency. This trend is more clearly visible as the stellar mass increases. This is evident from Fig.~\ref{fig:near_f_zoom_m198}, where we plot the comparison between the tidal responses in the current work and the result from Ref.~\cite{Andersson:2025iyd} near the f-mode for a star with mass $1.98M_\odot$, and radius $R=12.59$~km. The dashed vertical line shows the true value of the real part of the f-mode frequency. It is clear that the earlier result in Ref.~\cite{Andersson:2025iyd} has a larger offset (of around $0.1\%$) towards higher frequencies, whereas $k_2^{\rm EFT}(\omega)$ exhibits resonant behaviour very close to the expected point. This suggests an improved capturing of tidal resonances for more massive neutron stars.

\subsection{The low-frequency behaviour}

The low-frequency behaviour of the tidal response provides a key sanity check of any dynamical tide model. Given this, it makes sense to explore how our results approach the known results for the static Love number~\cite{Hinderer_2008}. It was already demonstrated in~\cite{Saketh:2024juq} that the scattering problem can be matched consistently in the low-frequency limit to obtain the expected static tidal response. Nevertheless, the comparison we want to make is  not straightforward because the interior neutron-star problem is known to be tricky at low frequencies. The  problem is associated with terms in the solutions that scale with inverse powers of the frequency~\cite{Andersson:2025iyd,HegadeKR:2024agt,Saketh:2024juq}. This makes it difficult to approach the limit numerically. Nevertheless, we can establish that the tidal response smoothly approaches the expected result. We can also make a direct comparison to other results in this regime, like those from ~\cite{Andersson:2025iyd}. 

\begin{figure}[h]
    \includegraphics[width=\linewidth]{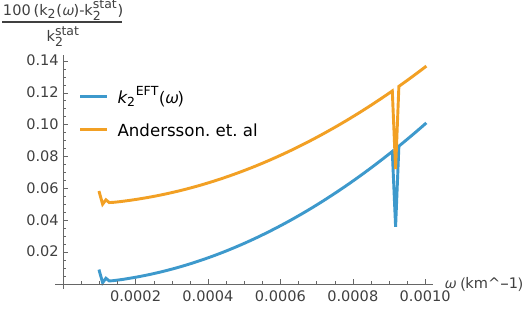}
    \includegraphics[width=\linewidth]{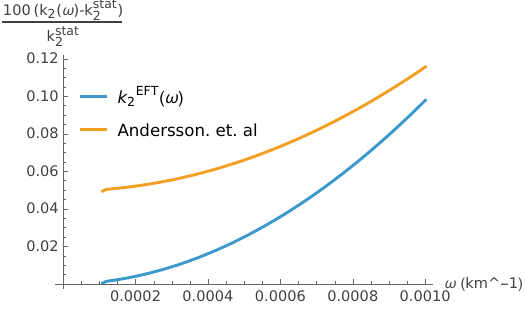}
    \caption{Low-frequency behaviour of the tidal response for the BSk22 equation of state and two stellar models. The top panel shows results for a star with $M=1.4M_\odot$ and $R=13.04~\mathrm{km}$ while the bottom panel represents a model with  $M=1.98M_\odot$ and $R=12.59~\mathrm{km}$. The results show that the new model has significantly improved on the systematic error discussed in~\cite{Andersson:2025iyd}. It is also worth noting the slight wiggles at the lowest frequencies in the top panel. These indicate the low-frequency region where numerically solving the neutron-star perturbation equations is difficult.  }
\label{fig:threeplots}
\end{figure}

In Fig.~\ref{fig:threeplots} we show the percent deviation from the static Love number for the tidal response obtained from Eq.~\eqref{eq:dynamical_tides} alongside   that from Ref.~\cite{Andersson:2025iyd}. This figure demonstrates the main improvement brought by our new model, which has significantly decreased the systematic deviation noted in~\cite{Andersson:2025iyd}.

\subsection{Estimating the imaginary part of the f-mode}

Another useful check of the relationship between the tidal scattering phase and the tidal response given in Eq.~\eqref{eq:EFTphasexp} may be performed by assuming that a single mode dominates the tidal response in the vicinity of resonance and inferring the imaginary part of the associated resonant mode frequency. In essence, the strategy extends the ``near-zone boundary condition'' proposal from Lindblom et al.~\cite{Lindblom_1997}. They demonstrated how neutron-star mode frequencies can be calculated by incorporating the outgoing-wave boundary condition in the weak-field near zone (essentially at the surface of the star). This led to an accurate calculation of the f-mode oscillation frequency, but the imaginary part (the damping rate)  could only be obtained with $ \sim 10\%$ accuracy~\cite{Andersson:2025iyd}.  

Specifically, neutron star oscillation modes correspond to frequencies for which only outgoing radiation is present at infinity. Since the exponential of the scattering phase is closely related to the ratio of the coefficients of outgoing and incoming waves, as shown in Eq.~\eqref{eq:SPT_phase}, it is expected to diverge at each mode frequency ($\omega=\omega_{\rm qnm}$). Using the relation
\begin{equation}
    \frac{A^{\rm out}_{\ell\omega}}{A^{\rm in}_{\ell\omega}}
\approx
(-1)^{\ell+1}
e^{2i\delta_{\ell\omega}^{pp}}
e^{2i\delta_{\ell\omega}^{\rm tidal}},
\end{equation}
and noting that the point-particle contribution to the phase does not contain information about the stellar matter properties, the neutron-star modes are expected to be associated with divergences in $e^{2i\delta_{\ell\omega}^{\rm tidal}}$. Making use of Eq.~\eqref{eq:EFTphasexp}, this leads to the condition
\begin{multline}
%\begin{split}
        1 = - i (R\omega_{\rm qm})^5\frac{F(\omega_{\rm qm})}{160 m_\mathrm{pl}^2 \pi}(1+2\pi M\omega_{\rm qm}) \\ =  i (R\omega_{\rm qm})^5 \frac{4 k_2(\omega_{\rm qm})}{15}(1+2\pi M\omega_{\rm qm}).
%\end{split}
\label{eq: qnmpole}
\end{multline}
Since $k_2(\omega)$ is real, the above relation cannot be satisfied for real frequencies, indicating the need for a complex solution, as expected given that the modes should be damped by gravitational-wave emission.

An approximate solution for the imaginary part of the dominant f-mode may be obtained by assuming that the f-mode dominates the tidal response in the near-resonance region. To do this, we make the  ansatz
\begin{equation}
    k_2(\omega)=\frac{A}
    {1-(\omega/\omega_{0})^2},
    \label{eq: one-mode}
\end{equation}
where $\omega_0$ and $A$ can be extracted by fitting $k_2(\omega)$ close to the f-mode. As seen from Fig.~\ref{fig:near_f_zoom}, $\omega_0$ is very close to the real part of the f-mode frequency. Substituting Eq.~\eqref{eq: one-mode} into Eq.~\eqref{eq: qnmpole} and expanding around $\omega_{\rm qnm}\approx\omega_0+\delta\omega$, we obtain
\begin{equation}
    \delta \omega \approx -i\frac{2\omega_0}{15}A(R\omega_0)^5(1+2\pi M\omega_0) + \mc{O}(A^2).   \label{eq:imf}
\end{equation}
The leading-order correction to the mode frequency is negative and imaginary, as expected. Implementing this strategy, we have evaluated this expression for the BSk equation of state and for three stellar models, including the two previously considered. The results are compared with the known values of the imaginary part of the f-mode in Table.~\ref{tab:omega}. As discussed earlier, we identify $\omega_0$ with the real part of the f-mode frequency and we already know from comparisons with Ref~\cite{Andersson:2025iyd} in Figs.~\ref{fig:four_panel_near_modes}, \ref{fig:near_f_zoom}, that this is accurately determined from our calculation.

\begin{widetext}
\begin{center}
\centering
\begin{table}[h!]
\centering
\renewcommand{\arraystretch}{1.15}
\setlength{\tabcolsep}{7pt}

\begin{tabular}{c c c c c c c c}
\hline\hline
$M/M_{\odot}$ & $R~(\mathrm{km})$ &
$\Re(\omega_f)~(\mathrm{km}^{-1})$ &
$-\Im(\omega_f)~(\mathrm{km}^{-1})$ & A & $\omega_0$ &
$-\Im(\delta\omega)~(\mathrm{km}^{-1})$ &
$\begin{array}{c}
100\times\left|
\dfrac{i\Im(\omega_f)-\delta\omega}
{i\Im(\omega_f)}
\right|
\end{array}$ \\
\hline

 1.4 & 13.04 & 0.03447 &
$1.2646\times10^{-5}$ & 0.1036 & 0.03447 &
$1.2646\times10^{-5}$ &
$\le 0.002\%$ \\[4pt]

 1.6 & 12.98 & 0.03628 &
$1.58536\times10^{-5}$ & 0.09187& 0.03628 &
$1.58537\times10^{-5}$ &
$\le 0.001\%$ \\[4pt]

 1.98 & 12.59 & 0.04034 &
$2.14998\times10^{-5}$ & 0.06781 & 0.4034 &
$2.15\times 10^{-5}$
&
$\leq 0.001\%$
\\

\hline\hline
\end{tabular}

\caption{Imaginary part of the f-mode for stellar models determined from the BSk22 equation of state as predicted by the pole in Eq.~\eqref{eq:EFTphasexp} using the numerically fitted analytical ansatz in Eq.~\eqref{eq: one-mode}. Note that the disagreement is essentially negligible and is thus a significant improvement over previous results in Ref.~\cite{Andersson:2025iyd} using the method advanced in Ref.~\cite{Lindblom_1997}}

\label{tab:omega}
\end{table}
\end{center}
\end{widetext}

We find that the predicted imaginary parts of the calculated f-modes agree exceptionally well with the known values. In fact, they represent a dramatic improvement of the results reported in ~\cite{Andersson:2025iyd}. This shows the robustness and physical validity of our method. 

It is relevant to add a couple of comments on the numerical results provided in table~\ref{tab:omega}. The calculation is subject to several sources of uncertainty, including the finite resolution and interpolation of the equation of state table, as well as the numerical accuracy of the solvers used for the stellar interior and perturbation equations. However, these effects are common to both determinations of the mode frequency, namely the direct output of the mode solver and the estimate obtained from Eq.~\eqref{eq:imf}, since they use the same interior solution for the background of the star as well as the numerical solutions for the metric and matter perturbations. Consequently, any systematic bias arising from these sources is expected to affect the two methods in a similar manner and should therefore have only a limited impact on the comparison.

\subsection{Discussion of results}

Before concluding, it is important to emphasize that the ultimate objective of any definition of the relativistic tidal response is not merely to characterize the response itself, but to incorporate tidal effects into binary dynamics and, ultimately, into robust gravitational waveform models. Consequently, the merits of a given approach to dynamical tides should be assessed not only by the form of the resulting tidal response, but also by the theoretical and computational framework that underlies it.

A key advantage of the present formulation is that it is directly connected to the underlying dynamics of the compact object within the (now standard) PN-EFT frameworks. The tidal response derived here provides a systematic prescription for incorporating tidal dynamics into the worldline action of Eq.~\eqref{eq:action2}, from which the influence on the motion of a compact object subject to an external field can be computed in a consistent manner. Following earlier works such as Ref.~\cite{Mandal:2023hqa}, it can also be incorporated straightforwardly into a binary Hamiltonian within the PN formalism. Moreover, our approach provides a systematic treatment of the exterior spacetime and yields a manifestly gauge-invariant definition of the tidal response.

While several previous works have formulated the relativistic tidal response within a worldline EFT framework~\cite{Saketh:2022xjb,Saketh:2023bul,Mandal:2023hqa,Steinhoff:2016rfi,Goldberger:2005cd,Goldberger:2004jt,Saketh:2024juq}, they typically either restrict attention to the low-frequency adiabatic regime, thereby excluding resonant phenomena, or model the tidal quadrupole as a single harmonic oscillator associated with the fundamental mode.\footnote{It is important to mention Ref.~\cite{Martinez-Rodriguez:2026omk} here which systematically investigated the validity of modeling the tidal response as a sum of harmonic-oscillator responses in an idealized perfect fluid neutron star.} In contrast, our construction avoids both the adiabatic approximation and the one-mode assumption. Instead, it is rigorously matched to stellar perturbation theory in a gauge-invariant manner, enabling the full dynamical tidal response of the compact object to be captured within the EFT.

An equally important advance of the present framework, building on developments in Refs.~\cite{Caron-Huot:2025tlq,Ivanov:2026icp,Saketh:2023bul}, is its systematic improvability. The expression for the tidal response in Eq.~\eqref{eq:dynamical_tides} can be refined by computing additional diagrams at higher orders in $M$, thereby incorporating progressively more of the exterior spacetime dynamics. Indeed, several relevant  ingredients are already available and can be incorporated directly into the formalism, including some higher-order loop corrections~\cite{Goldberger:2009qd} and resummed tail effects~\cite{Goldberger:2009qd,Chang:2026eti}. Such flexibility is particularly important for the coming era of precision gravitational-wave astronomy, where efficient, systematic, and internally consistent treatments of the binary problem are essential for reducing waveform systematics and fully exploiting the capabilities of future detectors.

The fact that our approach correctly reproduces resonant behaviour~(see Fig.~\ref{fig:four_panel_near_modes}, Fig.~\ref{fig:near_f_zoom}, and Fig.~\ref{fig:near_f_zoom_m198}), possesses a consistent low-frequency limit without introducing systematic errors (see Fig.~\ref{fig:threeplots}), and enables accurate determination of the imaginary parts of the f-mode frequencies (see Table.~\ref{tab:omega}) all lend support to the validity of the method. However, these phenomenological successes should be viewed as consequences of a broader achievement: the construction of a gauge-invariant, systematically improvable EFT framework for relativistic dynamical tides in neutron stars that is directly suited for applications to compact-binary dynamics and waveform modeling.

\section{Conclusion and future work}
\label{secV:CONC}

We have developed a new frequency-dependent model for the tidal response of a neutron star, as defined within the WEFT framework, by matching to results from relativistic stellar perturbation theory, explicitly without a strict low-frequency expansion. In particular, we exploited the presence of a hierarchy of scales in a neutron star, ($\omega_g<\omega_f\ll M^{-1}$), to obtain a dynamical tidal response that can capture resonant features (where $\omega\sim \omega_{g,f}$), while still working within an EFT framework with a post-Newtonian/Minkowskian expansion in the parameter $M\omega$. Our proof-of-principle results show that the dynamical tidal response, including the regime near resonant frequencies, can be consistently determined up to corrections of order $(M\omega)^2$ relative to the leading contribution. Such corrections are generally negligible during binary inspiral, which is relevant at 3PN order with respect to the leading order tidal effects. Furthermore, our framework can be systematically extended to incorporate higher-order corrections in $M\omega$ as additional EFT results (i.e., newly computed diagrams) become available. Consequently, the framework is designed to naturally accommodate both current and future developments within WEFT.

Our model provides an analytical expression for the excess conservative dynamical tidal response of a neutron star with respect to that of a black hole, Eq.~\eqref{eq:dynamical_tides}, which reduces to the well-known result for the static Love number in the appropriate low-frequency limit~\cite{Saketh:2024juq,Hinderer_2008}. Comparing our results to an earlier expression obtained in Ref.~\cite{Andersson:2025iyd}, we demonstrated that the two models lead to qualitatively similar behavior, including the expected features near resonant frequencies. However, the tidal response derived in this work is significantly more accurate for both low-frequencies~(see Fig.~\ref{fig:threeplots}) and near resonant mode frequencies (as evidenced by the accurate determination of the imaginary part of the fundamental mode in Table.~\ref{tab:omega}). While we have ignored the dynamical tidal response of the black hole, the resulting error is expected to be negligible, as it is suppressed by powers of both the PN expansion parameter and the compactness, scaling as $\sim (M\omega)^2 (R\omega)^5$. Furthermore, as noted previously, several waveform models use the black hole binary waveform as the baseline and incorporate neutron star effects as perturbative corrections. From this practical perspective, our approach should provide the appropriate prescription for incorporating neutron star dynamical tides~\cite{Gralla:2017djj}.

Although the numerical differences between the tidal response obtained in this work and those derived in previous work are generally relatively small, it is important to emphasize that the way in which our tidal response enters the binary dynamics is particularly transparent. Unlike most other approaches, which tend to define the tidal response through the ratio of growing and decaying metric perturbation modes in the a region near the star's surface or through other characteristics of the perturbed stellar metric, our definition is based solely on a quantity appearing in the worldline action in the EFT. This quantity directly encodes the dynamical properties of the neutron star and can be  straightforwardly incorporated into the computation of the corresponding PN Hamiltonian, e.g. by following the analysis from for example Ref.~\cite{Mandal:2023hqa}. Furthermore, by relating the tidal response to an observable quantity, namely the tidal scattering phase, we avoid ambiguities associated with coordinate freedom.

This work can be extended in several directions.
On the EFT side, a more careful analysis of the linear order in mass result in Eq.~\eqref{eq:treeMsummed} is required, which for now has simply been "guessed" based on the lowest-order results and the unitarity conditions. A straightforward extension after that would be to incorporate effects at order $(M\omega)^2$ and higher, which have already been partially calculated in other contexts~\cite{Goldberger:2009qd, Saketh:2023bul, Mandal:2023hqa,Jakobsen:2023pvx,Edison:2024owb}. The next order, $(M\omega)^2$, is also where the EFT tidal response is known to require renormalization due to loop divergences, leading to a logarithmic frequency dependence in related observables~\cite{Mandal:2023hqa,Saketh:2023bul}. Another important direction would be the inclusion of viscous fluid dissipation within the stellar interior,  recently shown to have significant implications for the presence of exotic matter in neutron-star interiors.~\cite{Ghosh:2023vrx,Ghosh:2025wfx}. Such effects were previously studied within a low-frequency approximation~\cite{Saketh:2024juq}, but are expected to become more significant near resonances and should therefore be incorporated into the dynamical tidal response. Finally, one can instead focus on using the WEFT to derive a two-body Hamiltonian following previous works such as in Ref.~\cite{Mandal:2023hqa}, but without assuming an oscillator model or taking the adiabatic limit, to explicitly show how the tidal response derived in this work enters binary dynamics.     

On the side of stellar perturbation theory, one may explore a  range of neutron-star equations of state, including exotic scenarios like dark-matter admixed models. Arguably, the most interesting technical development would be to incorporate spin effects. This will require  modifications to both the perturbation theory and EFT frameworks, but also brings new physics into play. In particular, one would be able to quantify the effect of rotational frame dragging and the impact of tidal resonances associated with inertial modes (which rely on the gravito-magnetic coupling). We plan to explore some of these directions in future work.

\section*{acknowledgements}

We thank Leonardo Gualtieri, Abhishek Hegade, Godwin Martin, Soumodeep Mitra, Raj Patil, Eric Poisson, Jan Steinhoff, and Zihan Zhou for discussions and feedback. The research of M.V.S.S. is supported by the National Post-Doctoral Fellowship (PDF/2025/004764), ANRF, Government of India. N.A. and S.G. gratefully acknowledge support from the STFC via Grant
No. ST/Y00082X/1.

\end{document}